\documentclass[11pt,letterpaper]{article}
\pdfoutput=1
\usepackage{jinstpub}
\usepackage{color}
\usepackage[table]{xcolor}
\usepackage{graphicx}

\usepackage{verbatim}
\usepackage{amsmath}
\usepackage{amssymb}
\usepackage{subfig}
\usepackage{url}
\usepackage{bbold}
\usepackage{slashed}
 \usepackage{url}
\usepackage{multirow}
\usepackage{threeparttable}
\usepackage{paralist}
\usepackage{bm}
\usepackage{hyperref}
\hypersetup{
    colorlinks=true,
    linkcolor=blue,
    filecolor=magenta,      
    urlcolor=cyan,
}

\DeclareRobustCommand{\Tab}[1]{Table~\ref{#1}}

\newcommand{\be}{\begin{equation}}
\newcommand{\ee}{\end{equation}}

\newcommand{\tev}{\mathrm{TeV}}

\newcommand{\pt}{p_{\mathrm{T}}}

\newcommand{\hlsfml}{{\tt{hls4ml}}}

\renewcommand{\vec}[1]{\bm{#1}} % ISO complying version?
\newcommand{\matr}[1]{\bm{#1}}     % ISO complying version

\newcommand{\norm}[1]{\|#1\|}

\begin{document}
%\title{Deep Neural Networks in FPGAs for Trigger and DAQ}
%\title{Deep neural networks in FPGAs for real-time particle physics applications}
\title{Fast inference of deep neural networks in FPGAs for particle physics}

\author[a]{Javier Duarte}
\author[b]{, Song Han}
\author[b]{, Philip Harris}
\author[a]{, Sergo Jindariani}
\author[c]{, Edward Kreinar}
\author[a]{, Benjamin Kreis}
\author[d]{, Jennifer Ngadiuba}
\author[d]{, Maurizio Pierini}
\author[a]{, Ryan Rivera}
\author[a]{, Nhan Tran}
\author[e]{, Zhenbin Wu}

\affiliation[a]{Fermi National Accelerator Laboratory, Batavia, IL 60510, USA}
\affiliation[b]{Massachusetts Institute of Technology, Cambridge, MA 02139, USA}
\affiliation[c]{HawkEye360, Herndon, VA 20170, USA}
\affiliation[d]{CERN, CH-1211 Geneva 23, Switzerland}
\affiliation[e]{University of Illinois at Chicago, Chicago, IL 60607, USA}

%\emailAdd{ntran@fnal.gov}
\emailAdd{hls4ml.help@gmail.com}

\abstract{
Recent results at the Large Hadron Collider (LHC) have pointed to enhanced physics capabilities 
through the improvement of the real-time event processing techniques. 
Machine learning methods are ubiquitous and have proven to be very powerful in LHC physics, and particle physics as a whole.  
However, exploration of the use of such techniques in low-latency, low-power FPGA (Field Programmable Gate Array) hardware has only just begun.
FPGA-based trigger and data acquisition systems have extremely low, sub-microsecond latency requirements that are unique to particle physics.
We present a case study for neural network inference in FPGAs focusing on a 
classifier for jet substructure which would enable, among many other physics scenarios, searches for new dark sector particles
and novel measurements of the Higgs boson. 
While we focus on a specific example, the lessons are far-reaching.
A companion compiler package for this work is developed based on High-Level Synthesis (HLS) called \hlsfml~to build machine learning models in FPGAs.
%We develop a package based on High-Level Synthesis (HLS) called \hlsfml~to build machine learning models in FPGAs.
The use of HLS increases accessibility across a broad user community and allows for a drastic decrease in firmware development time.
We map out FPGA resource usage and latency versus neural network hyperparameters 
to identify the problems in particle physics that would benefit from performing neural network inference with FPGAs. 
For our example jet substructure model, we fit well within the available resources of modern FPGAs with a latency on the scale of 100~ns.
}

%\preprint{
\begin{flushright}
  FERMILAB-PUB-18-089-E 
\end{flushright}
%}

\maketitle

%%%%%%%%%%%%%%%%%%%%%%%%%%%%%%%%%%%%%%%%%%%%%%%%%%%%%%%%%%%%%%%%%%%%%%%%%%%%%%%%%%%%%%%%%%%%%%%%
% I N T R O D U C T I O N
%%%%%%%%%%%%%%%%%%%%%%%%%%%%%%%%%%%%%%%%%%%%%%%%%%%%%%%%%%%%%%%%%%%%%%%%%%%%%%%%%%%%%%%%%%%%%%%%
%\clearpage
\section{Introduction}
\label{sec:introduction}
Over the last several decades, as physicists have pursued a deeper understanding of particle physics phenomena, particle detectors have been made larger, more granular, and capable of processing data at ever increasing rates. This has led to a dramatic increase in data volumes that need to be efficiently analyzed in real time to reconstruct and filter events of interest.
%Deep learning
Machine learning (ML) methods deployed in the final stages of data processing have been demonstrated to be extremely effective in many different tasks across particle physics. Thus far, their use in real-time selection hardware, based on Field Programmable Gate Arrays (FPGAs), has been limited due to implementation complexity and FPGA resource demands. In this study, we explore the implementation of neural networks in FPGAs, mapping out resource usage and latency for various deep neural network architectures and hyperparameters, demonstrating the feasibility of deep learning techniques in very low-latency (sub-microsecond) FPGA applications.

The Large Hadron Collider (LHC) at CERN serves as a perfect example.
The LHC is the world's highest energy particle accelerator,
operating at the highest data rates ever achieved. %PH:added data to rates% 
Its goal is to understand the very basic laws of nature and the building blocks of the universe. 
In the first run of data-taking that concluded in 2012, the ground-breaking highlight was 
the discovery of the Higgs boson~\cite{Aad:2012tfa,Chatrchyan:2012xdj}.
The current data-taking campaigns are devoted to a full characterization of the Higgs-boson properties
and to the search for physics phenomena beyond the standard model of particle physics, including the search for dark matter.

Due to the extreme frequency of 40~MHz at which proton bunches collide at the LHC, data rates at the two multipurpose experiments, CMS~\cite{Chatrchyan:2008aa} and ATLAS~\cite{Aad:2008zzm}, are of the order of hundreds of terabytes per second.  
With such high data rates, there are challenges for both real time and offline processing of collision events.  
The task of the real-time processing is to filter events to reduce data rates to manageable levels for offline processing is called triggering.
It is typically performed in multiple stages~\cite{Bernius:2017kyk,Tosi:2017hjj}; two stages are used in the CMS detector.
Because of the extreme input data rates and size of the data buffers, 
the first stage, Level-1 (L1), of data processing typically uses custom hardware with ASICs or, increasingly, FPGAs, to handle the initial data rate
using pipelined algorithms with latencies of hundreds of nanoseconds totalling microseconds.
The second stage of triggering, High Level Trigger (HLT), uses commercial CPUs to process the filtered data in software with longer latencies on the timescale of hundreds of milliseconds in total. 

Machine learning methods, most recently deep learning, 
have a wide range of applications in event processing at the LHC~\cite{Sirunyan:2018ouh,Aaboud:2018wps,Sirunyan:2017ezt,Aaboud:2018xwy,1742-6596-898-6-062009,Sirunyan:2018hoz,Aaboud:2018urx}, from lower level energy cluster calibration and regression to high level physics object classification, e.g. jet tagging with substructure information, and physics analyses.
%While ML methods are typically employed in offline analysis, recent results at the Large Hadron Collider (LHC) have pointed to enhanced physics capabilities through the improvement of real-time event processing.
More sophisticated ML algorithms in the trigger will allow LHC experiments to preserve potential new physics signatures such as those related to the Higgs, dark matter, and hidden sectors~\cite{Gershtein:2017tsv} that would otherwise be lost.  
This can be achieved through overall more performant trigger algorithms or fast data analysis techniques such as data scouting or trigger level analysis~\cite{Khachatryan:2016ecr,Aaboud:2018fzt,Sirunyan:2016iap}.
In this study, we explore the implementation of neural network inference in FPGAs by
mapping out resource usage and latency for various network architectures and hyperparameters.
The goal is to understand the range of possible neural network designs that may be implemented in FPGAs given the resource and latency constraints for different LHC event processing applications.
As a case study, we consider the case of using ML methods for jet substructure~\cite{Larkoski:2017jix} classification which, when employed in the trigger, enable searches for new dark sector particles and important measurements of the Higgs transverse momentum spectrum.  
The lessons learned in this case study are broadly applicable. 

An important complementary result of this work is the companion compiler, \hlsfml\footnote{The project can be accessed at \href{https://hls-fpga-machine-learning.github.io/hls4ml}{https://hls-fpga-machine-learning.github.io/hls4ml}}, which translates ML models from common open-source software packages such as {\tt{Keras}}~\cite{chollet2015keras} and {\tt{PyTorch}}~\cite{paszke2017automatic} into RTL (Register-Transfer Level) abstraction for FPGAs using High-Level Synthesis (HLS) tools, which have demonstrated considerable progress in recent years~\cite{7082747}.
There are many types of HLS, and our particular package and the results of this study are based on Vivado HLS {\tt 2017.2}~\cite{vivadohls} though the general techniques can be used in other applications with other FPGAs.

In high energy physics, engineering support with long development cycles is required to translate physics-motivated data processing algorithms into firmware.  
However, engineering is a scarce and valuable resource.
The \hlsfml~tool allows physicists to rapidly prototype ML algorithms for both firmware feasibility and physics performance without extensive Verilog/VHDL experience, thus greatly decreasing the time for algorithm development cycles while preserving engineering resources.
We focus on the task of the FPGA-based triggers of the ATLAS and CMS experiments with algorithm latencies in the microsecond range, fully pipelined to handle the 40~MHz LHC collision rate.  For this task, solutions with either CPUs or GPUs are not possible due to the severe time limitation imposed. 
Such latencies are unique to LHC trigger challenges, and therefore few general tools exist for this application.  
Nonetheless, 
the \hlsfml~package is a general purpose tool and is designed to serve a broad range of applicatons in particle physics and beyond, from trigger and data acquisition tasks (DAQ) to longer latency trigger tasks (milliseconds) and CPU-FPGA co-processor hardware.

The rest of the paper is organized as follows. 
In Sec.~\ref{sec:hls4ml}, we describe the essential concepts in implementing neural networks in FPGAs for trigger and DAQ.  
This includes a case study for developing a jet substructure ML algorithm for FPGA implementation.
In Sec.~\ref{sec:results}, we detail the HLS synthesis for various neural network architectures and hyperparameters and the resulting implementation on an FPGA.
Finally, we summarize our findings, detail follow-up studies, and discuss potential broader applications in physics and other fields in Sec.~\ref{sec:outlook}.

\subsection{Related Work}

The inference of neural networks in FPGAs is a rapidly developing and high interest field.
There exists an extensive literature on the topic.  
However, as it pertains to the particular task for particle physics where networks can be smaller but latency constraints are much more severe, this is the first dedicated general purpose study of this kind in the field.
Nonetheless, some ML techniques have been deployed in the LHC trigger already, including a first implementation of a boosted decision tree (BDT) for muon momentum measurement~\cite{Acosta:2290188}. An early attempt to deploy convolutional neural networks (CNNs) on FPGAs for particle physics was presented at NIPS~2017~\cite{Calafiura-NIPS}.
%Imperial BDT in logic??

We borrow inspiration from other works, including the RFNoC Neural Network Library~\cite{grcon}, on which \hlsfml~is based. An overview of existing toolflows for mapping CNNs on FPGAs is given in~\cite{2018arXiv180305900V}. Snowflake~\cite{snowflake} is a scalable and efficient CNN accelerator with models specified in {\tt{Torch}}~\cite{torch} and a single computation architecture (sequential IO) designed to perform at near-peak hardware utilization targeting Xilinx System-on-Chips (SoCs).  %, such as Zynq XC7Z045. 
Caffeine~\cite{7827589} is another CNN accelerator for {\tt{Caffe}}-specified models targeting Xilinx devices that support a SDAccel15 environment and a PCIe interface between the FPGA and a host. 
fpgaConvNet~\cite{Venieris_2017nips,venieris2017fpga,venieris2017fpl,venieris2016fccm} converts CNNs specified in {\tt{Caffe}}~\cite{jia2014caffe} or {\tt{Torch}} formats into generated Xilinx Vivado HLS code with a streaming architecture (parallel IO). 
FP-DNN (Field Programmable Deep Neural Networks)~\cite{7966671} is a framework that takes {\tt{TensorFlow}}~\cite{tensorflow2015-whitepaper}-described DNNs (CNNs, LSTM-RNNs~\cite{LSTM}, and Residual Nets) as input, and generates the hardware implementations on FPGA boards with RTL-HLS hybrid templates. 
DNNWeaver~\cite{dnnweaver:micro16} is an open-source alternative, which also supports DNNs specified in {\tt{Caffe}} format and automatically generates the accelerator Verilog code using hand-optimized Verilog templates with a high degree of portability.
% (targeting SoCs and server-grade FPGAs from both Xilinx and Intel, including the Xilinx Zynq XC7Z020 SoC and the larger Intel Stratix V GSD5 and Arria 10 GX115).

The physics problem we take as a benchmark for our discussion is substructure-based jet tagging, on which there is a rich literature of deep-learning applications~\cite{deOliveira:2015xxd,Guest:2016iqz,Macaluso:2018tck,Datta:2017lxt,Butter:2017cot,Kasieczka:2017nvn,Komiske:2016rsd,Schwartzman:2016jqu}. 
In this context, there were attempts to use CNN, RNNs, as well as physics-inspired network architectures. 
Jets have been represented as grayscale images, RGB images, sequences of particles, or a set of physics-inspired high-level features, as we do in this study.

% %%%%%%%%%%%%%%%%%%%%%%%%%%%%%%%%%%%%%%%%%%%%%%%%%%%%%%%%%%%%%%%%%%%%%%%%%%%%%%%%%%%%%%%%%%%%%%%%
% % H L S 4 M L
% %%%%%%%%%%%%%%%%%%%%%%%%%%%%%%%%%%%%%%%%%%%%%%%%%%%%%%%%%%%%%%%%%%%%%%%%%%%%%%%%%%%%%%%%%%%%%%%%
%\clearpage
\section{Building neural networks with \hlsfml}
\label{sec:hls4ml}
In this section, we give an overview of translating a given neural network model into a FPGA implementation using HLS.
We then detail a specific jet substructure case study, but the same concepts are applicable for a broad class of problems.  
We conclude this section by discussing how to create an efficient and
optimal implementation of a neural network in terms of
performance, resource usage, and latency.

\subsection{\hlsfml~concept}
\label{sec:hls4ml-concept}
\begin{figure}[tbh!]
\begin{center}
\includegraphics[width=0.90\linewidth]{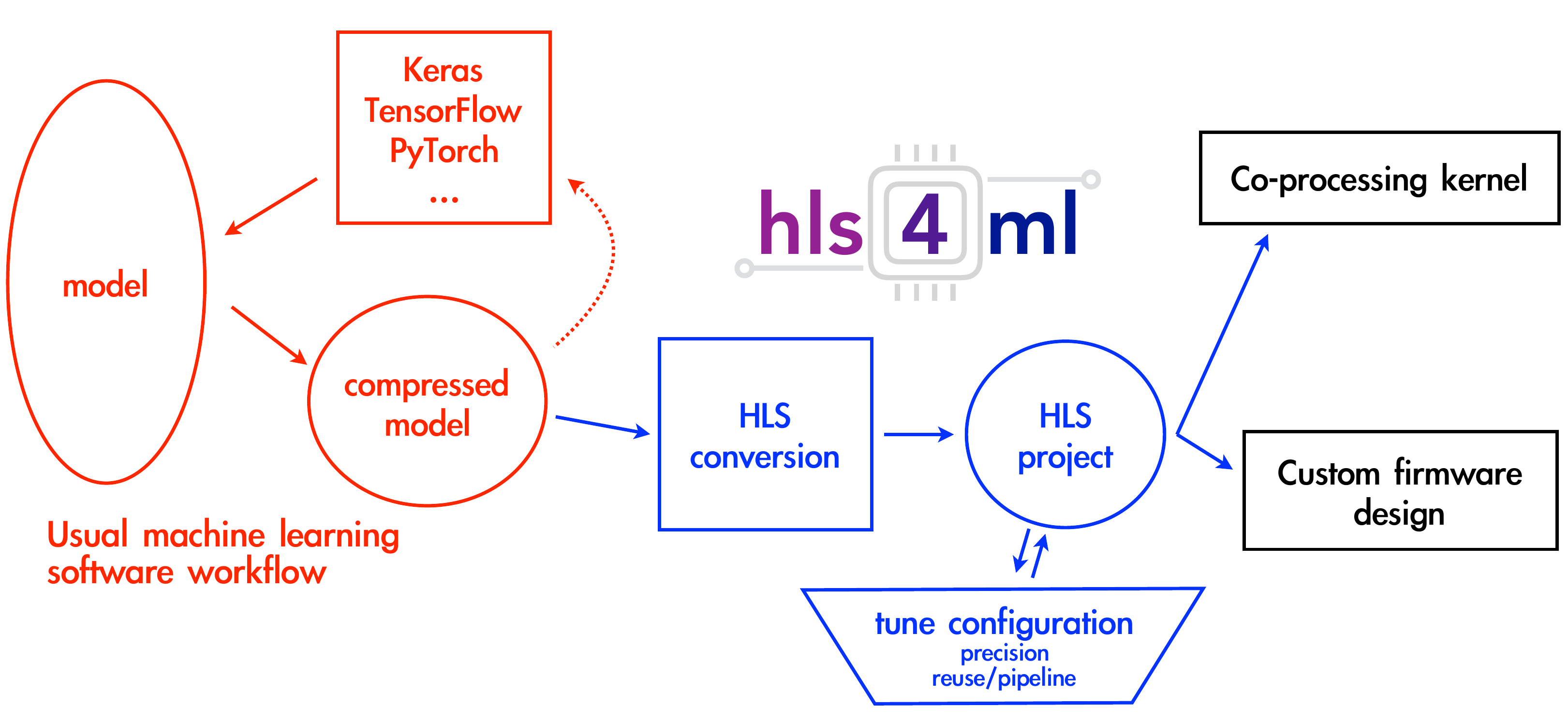}
\end{center}
\caption{A typical workflow to translate a model into a FPGA implementation using \hlsfml.}
\label{fig:flow}
\end{figure}

The task of automatically translating a trained neural network, specified by the model's architecture, weights, and biases, into HLS code is performed by the \hlsfml~package.
A schematic of a typical workflow is illustrated in Fig.~\ref{fig:flow}.

The part of the workflow illustrated in red indicates the usual software workflow required to design a neural network for a specific task.
This usual machine learning workflow, with tools such as {\tt Keras} and {\tt PyTorch}, involves a training step and possible compression steps (more discussion below in Sec.~\ref{sec:hls4ml-design}) before settling on a final model.
The blue section of the workflow is the task of \hlsfml, which translates a model into an HLS project that can be synthesized and implemented to run on an FPGA.

At a high level, FPGA algorithm design is unique from programming a CPU in that independent operations may run fully in parallel, allowing FPGAs to achieve trillions of operations per second at a relatively low power cost with respect to CPUs and GPUs. However, such operations consume dedicated resources onboard the FPGA and cannot be dynamically remapped while running. The challenge in creating an optimal FPGA implementation is to balance FPGA resource usage with achieving the latency and throughput goals of the target algorithm. Key metrics for an FPGA implementation include:

\begin{enumerate}
\item {\bf latency}, the total time (typically expressed in units of ``clocks'') required for a single iteration of the algorithm to complete.
\item {\bf initiation interval}, the number of clock cycles required before the algorithm may accept a new input. Initiation interval (often expressed as ``II'') is inversely proportional to the inference rate, or throughput; an initiation interval of 2 achieves half the throughput as an initiation interval of~1.  Consequently, data can be pipelined into the algorithm at the rate of the initiation interval.  
\item {\bf resource usage}, expressed as the following FPGA resource categories: onboard FPGA memory (BRAM), digital signal processing (arithmetic) blocks (DSPs), and registers and programmable logic (flip-flops, or FFs, and lookup tables, or LUTs).
\end{enumerate}

The \hlsfml~tool has a number of configurable parameters which can help the user explore and customize the space of latency, initiation interval, and resource usage tradeoffs for their application. Because every application is different, the goal of the \hlsfml~package is to empower the user to perform this optimization through automated neural network translation and FPGA design iteration. 
In practice, the time required to perform \hlsfml~translation of a neural netowrk is much shorter (minutes to hours) than a designing a specific neural network architecture for an FPGA, 
and may be used to rapidly prototype machine learning algorithms without dedicated engineering support for the FPGA implementation.
For physicists, this makes designing physics algorithms for the trigger or DAQ significantly more accessible and efficient, thus has the potential for the {\it "time to physics"} to be greatly reduced.

\begin{figure}[tbh!]
\begin{center}
\includegraphics[width=0.60\linewidth]{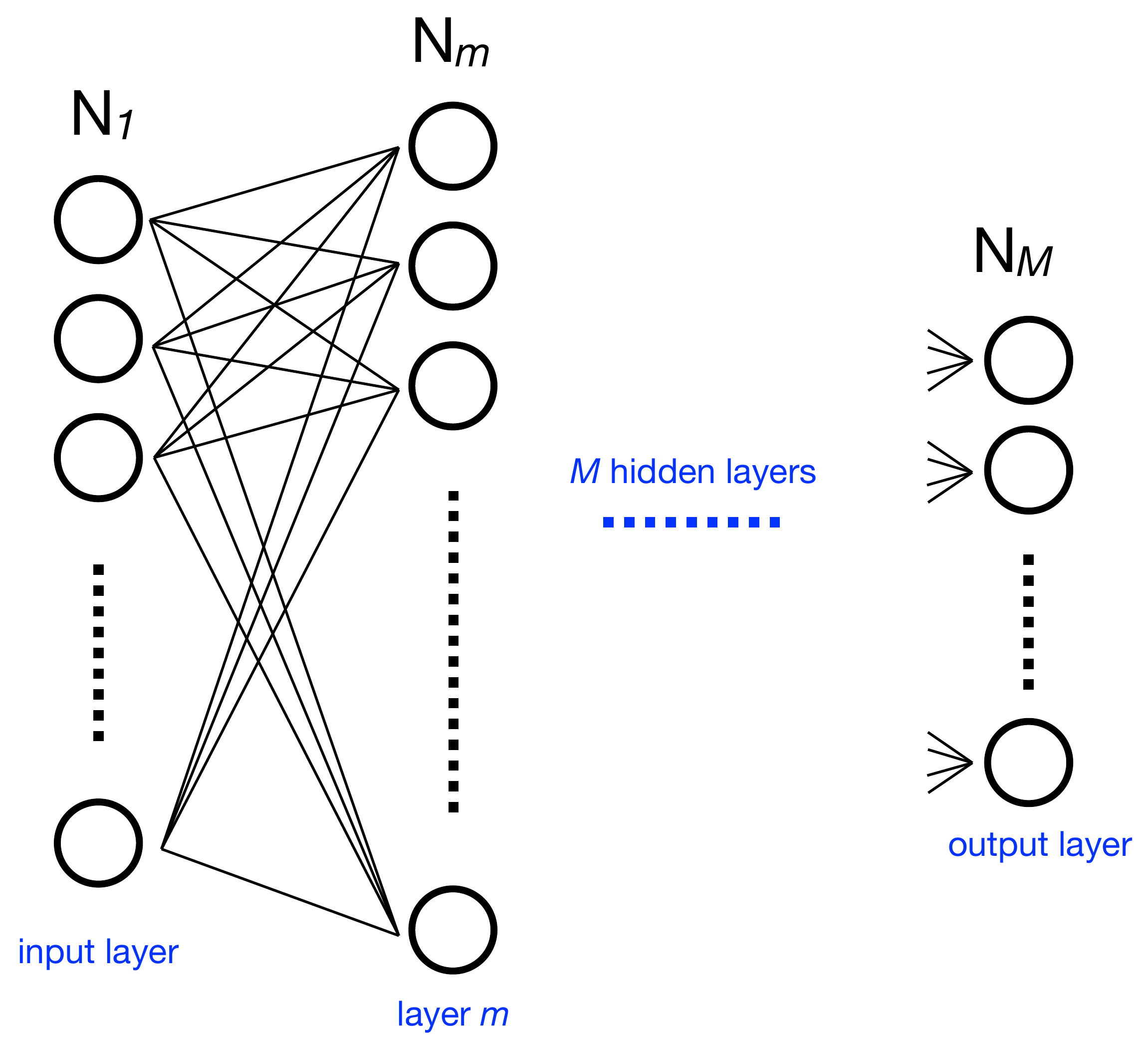}
\end{center}
\caption{A cartoon of a deep, fully connected neural network illustrating the description conventions used in the text}
\label{fig:dnn}
\end{figure}

We first introduce some terminology and concepts for the inference of deep, fully connected neural networks.
Consider the network illustrated in Fig.~\ref{fig:dnn} with $M$ layers, where each layer $m$ has $N_m$ neurons.  
The input layer has $N_1$ input neurons and the output layer has $N_M$ output neurons.
The vector of neuron output values at each layer are denoted by $\vec{x}_m$.
For the $m^\mathrm{th}$ fully connected layer ($m>1$), 
\begin{align}
\label{eq:vcalc}
\vec{x}_m &= g_m \left(\matr{W}_{m, m-1}  \vec{x}_{m-1} + \vec{b}_{m} \right)~,
\end{align}
where $\matr{W}_{m,m-1}$ is the matrix of weights between layers $m-1$ and $m$, $\vec{b}_m$ are the bias values, and $g_m$ is the activation function for layer $m$.
The size of matrix $\matr{W}_{m, m-1}$ is $N_{m} \times N_{m-1}$ and thus the number of multiplications required to compute the neuron values of layer $m$ is implicitly also $N_{m} \times N_{m-1}$.

In \hlsfml, the calculation of each layer $\vec{x}_m$ is performed independently and sequentially.
The inference is pipelined and accepts a new set of inputs after its initiation interval, as described above.
The total number of multiplications required to infer a given neural network is:
\begin{equation}
\label{eq:nmult}
N_{\rm multiplications} = \sum_{m=1}^{M} N_{m-1} \times N_{m}.
\end{equation}
Non-trivial activation functions, such as sigmoid, softmax, and hyperbolic tangent, are precomputed for a range of input values and stored in BRAMs.  
The ReLU activation function is implemented in programmable logic.
The effect of the neural network hyperparameters on the latency, throughput, and resource usage informs the optimal network implementation for any given application.

\subsection{Case study: jet substructure}
\label{sec:hls4ml-substructure}
Jets are collimated showers of particles that result
from the decay and hadronization of quarks $q$ and gluons $g$.  At the LHC, due to the high
collision energy, a particularly interesting jet signature emerges from 
overlapping quark-initiated showers produced in decays of heavy
standard model particles. For example, the $W$ and $Z$~bosons decay to
two quarks ($q\bar{q}$) 67\%-70\% of the time and the Higgs boson is predicted to
decay to two b-quarks ($b\bar{b}$) approximatly 58\% of the time.  The top quark decays to two
light quarks and a b-quark ($q\bar{q}b$).  It is the task of jet
substructure~\cite{Butterworth:2008iy,Larkoski:2017jix} to distinguish
the various radiation profiles of these jets from backgrounds
consisting mainly of quark ($u,d,c,s,b$) and gluon-initiated jets. % which subsequently ``shower'' to lower energy quarks and gluons. 
The tools of jet
substructure have been used to distinguish interesting jet 
signatures from backgrounds that have production
rates hundreds of times larger than the signal~\cite{Asquith:2018igt}.% that we wish to isolate.

Jet substructure at the LHC has been a particularly active field for machine learning techniques as jets contain $\mathcal{O}(100)$ particles whose properties and correlations may be exploited to identify physics signals.
The high dimensionality and highly correlated nature of the phase space makes this task an interesting testbed for machine learning techniques.
There are many studies that explore this possibility, both in experiment and theory~\cite{Larkoski:2017jix,Cogan:2014oua,Pearkes:2017hku,Louppe:2017ipp,deOliveira:2015xxd,Guest:2016iqz,Macaluso:2018tck,Datta:2017lxt,Butter:2017cot,Kasieczka:2017nvn,Komiske:2016rsd,Schwartzman:2016jqu,Butterworth:2008iy}.
For this reason, we choose to benchmark our FPGA studies using the jet substructure task. 

\begin{figure}[tbh!]
\begin{center}
\includegraphics[width=0.45\linewidth]{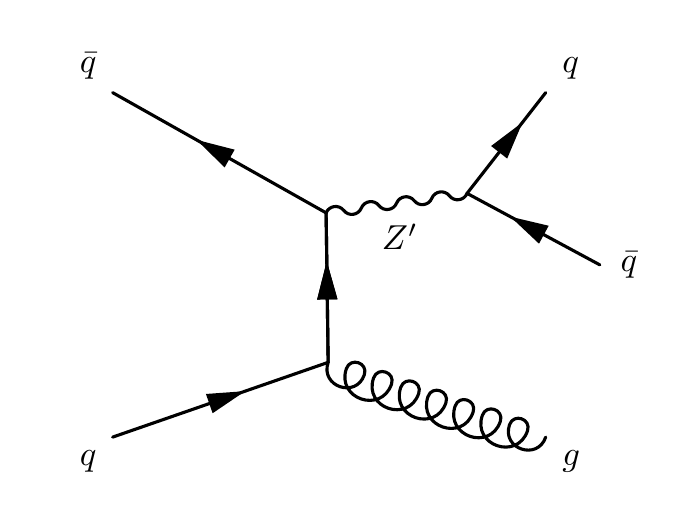}
\includegraphics[width=0.40\linewidth]{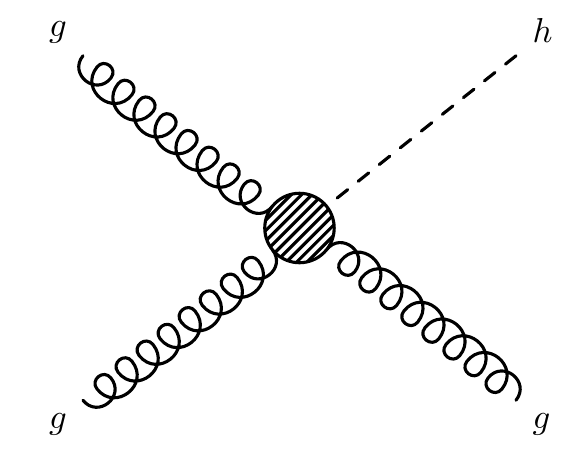}
\end{center}
\caption{Example Feynman diagrams of interesting physics signatures that would benefit from jet substructure algorithms in the hardware trigger.}
\label{fig:feyn}
\end{figure}

%For the trigger specifically, jet substructure techniques could be used to identify and preserve events containing interesting physics signatures that would typically be discarded.
We give two examples in Fig.~\ref{fig:feyn} where jet substructure techniques in the trigger can play an important role: low mass hidden hadronic resonances~\cite{Sirunyan:2017nvi} and boosted Higgs produced in gluon fusion~\cite{Sirunyan:2017dgc}.  
Both processes are overwhelmed by backgrounds and current trigger strategies would select only on the energy of the jet.
By introducing jet substructure techniques in the hardware trigger, we can further greatly reduced backgrounds and preserve significantly more of these types of signals in the future.  
There are many other physics signatures which could benefit from jet substructure in the trigger.
In this case study, we focus on the task of classifying a jet as either a quark ($q$), gluon ($g$), W boson ($W$), Z boson ($Z$), or top quark ($t$) jet
\footnote{The Higgs boson is not included in this study as its mass and substructure are quite similar to $W$ and $Z$ bosons and are otherwise difficult to distinguish in absence of track vertexing information, a situation common to current FPGA-based triggers.
}.

% --------------------------------------------------------------------------------------------------------------
\subsubsection*{Input generation and features}

Events are generated at $\sqrt{s} = 13~\tev$ for comparison to LHC performance.
Parton-level (unshowered quark) $W^+ W^-$, $ZZ$, $t\bar{t}$, $q\bar{q}$, and $gg$ events are first produced at leading-order using {\tt{MadGraph5\_aMC\_at\_NLO}}~\cite{Alwall:2014hca} (version 2.3.1) with the NNPDF23LO1 parton distribution functions (PDFs)~\cite{Ball:2012cx}.
To focus on a relatively narrow kinematic range, the transverse momenta of the partons and undecayed gauge bosons are generated in a window with energy spread given by $\delta \pt / \pt = 0.01$, centered at $1~\tev$.
These parton-level events are then decayed and showered in \textsc{Pythia8}~\cite{Sjostrand:2014zea} (version 8.212) with the Monash 2013 tune~\cite{Skands:2014pea}, including the contribution from the underlying event.
For each final state, 200,000 events are generated. 

To build a complete list of expert features, we implement a variety of jet recombination algorithms and substructure tools via the \textsc{FastJet~3.1.3} and \textsc{FastJet~contrib~1.027} packages~\cite{fastjet:1,fastjet:2}.
As a baseline, all jets are clustered using the anti-$k_{\rm{T}}$ algorithm~\cite{Cacciari:2008gp}, with a distance parameter of $R = 0.8$.
Even though the parton-level $\pt$ distribution is narrow, the jet $\pt$ spectrum is significantly broadened by kinematic recoil from the parton shower and energy migration in and out of the jet cone.
We apply a cut on the reconstructed jet $\pt$ to remove extreme events from the analysis, vetoing those outside a window of $0.8~\tev < \pt < 1.6~\tev$ for the $\pt = 1~\tev$ bin.

The jet substructure community has developed a wide variety of observables to identify the origin of a jet based on the structure of its radiation pattern.
In \Tab{tab:obslist}, we list all the observables used in this study~\cite{Adams:2015hiv,Larkoski:2014wba,Larkoski:2013eya,Moult:2016cvt}. 
A brief description of each of these variables is presented in Ref.~\cite{Coleman:2017fiq}.
These are used as expert-level inputs to a neural network classifier which is near optimal\footnote{More sophisticated approaches exist, but the goal of this study is not to achieve better performance than existing algorithms.  Instead, the goal is to examine the implementation of several effective neural network architectures in FPGAs.}.

\begin{table}[t]
\centering
\begin{tabular}{|c|c|}	\hline
\textbf{Observables} \\\hline\hline \hfill\\
$m_{\rm mMDT}$\\
$N_2^{\beta=1,2}$\\
$M_2^{\beta=1,2}$\\
$C_{1}^{\beta=0,1,2}$	\\
$C_{2}^{\beta=1,2}$ \\
$D_2^{\beta=1,2}$	\\
$D_2^{(\alpha,\beta)=(1,1),(1,2)}$	\\
$\sum z \log z$					\\
Multiplicity				\\ \hfill \\
\hline
\end{tabular}
\caption{A summary of the observables used in the analysis.}
\label{tab:obslist}
\end{table}

%%%%%%%%%%%%%%%%%%%%%%%%%%%%%%%%%%%%%%%%%%%%%%%%%%%%%%%%%%%%%%%%%%%%%%%%%%%%%%%%%%%%%%%%%%%%%%%%%%%%%%%%%%%%%%%%%%%%%%%%

\subsubsection*{Benchmark networks and floating point performance}
\label{ss:full3layerNN}

\begin{figure}[tbh!]
\begin{center}
\includegraphics[width=0.35\linewidth,clip=true,viewport=200 0 824 800]{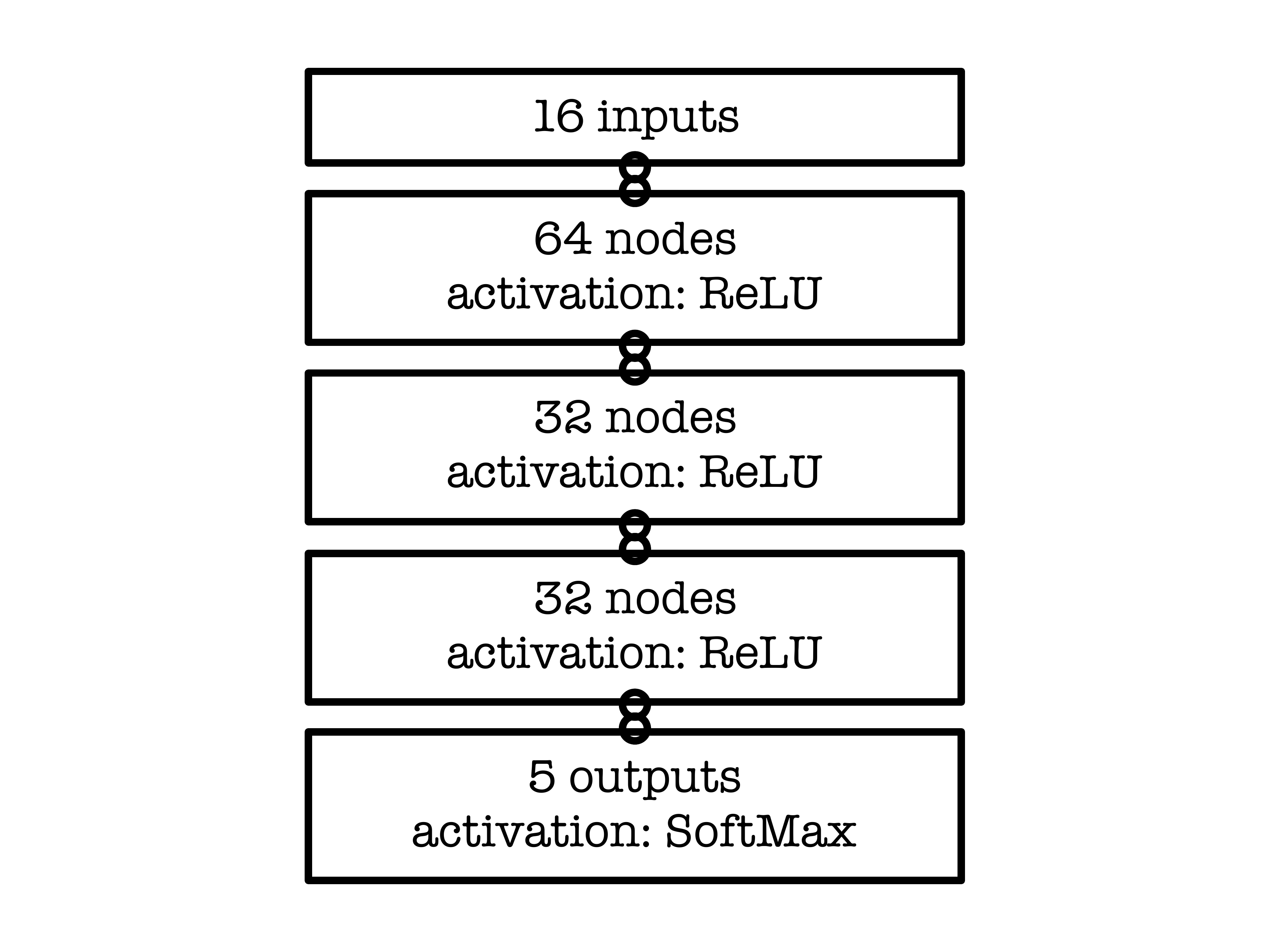}
\includegraphics[width=0.35\linewidth,clip=true,viewport=200 0 824 800]{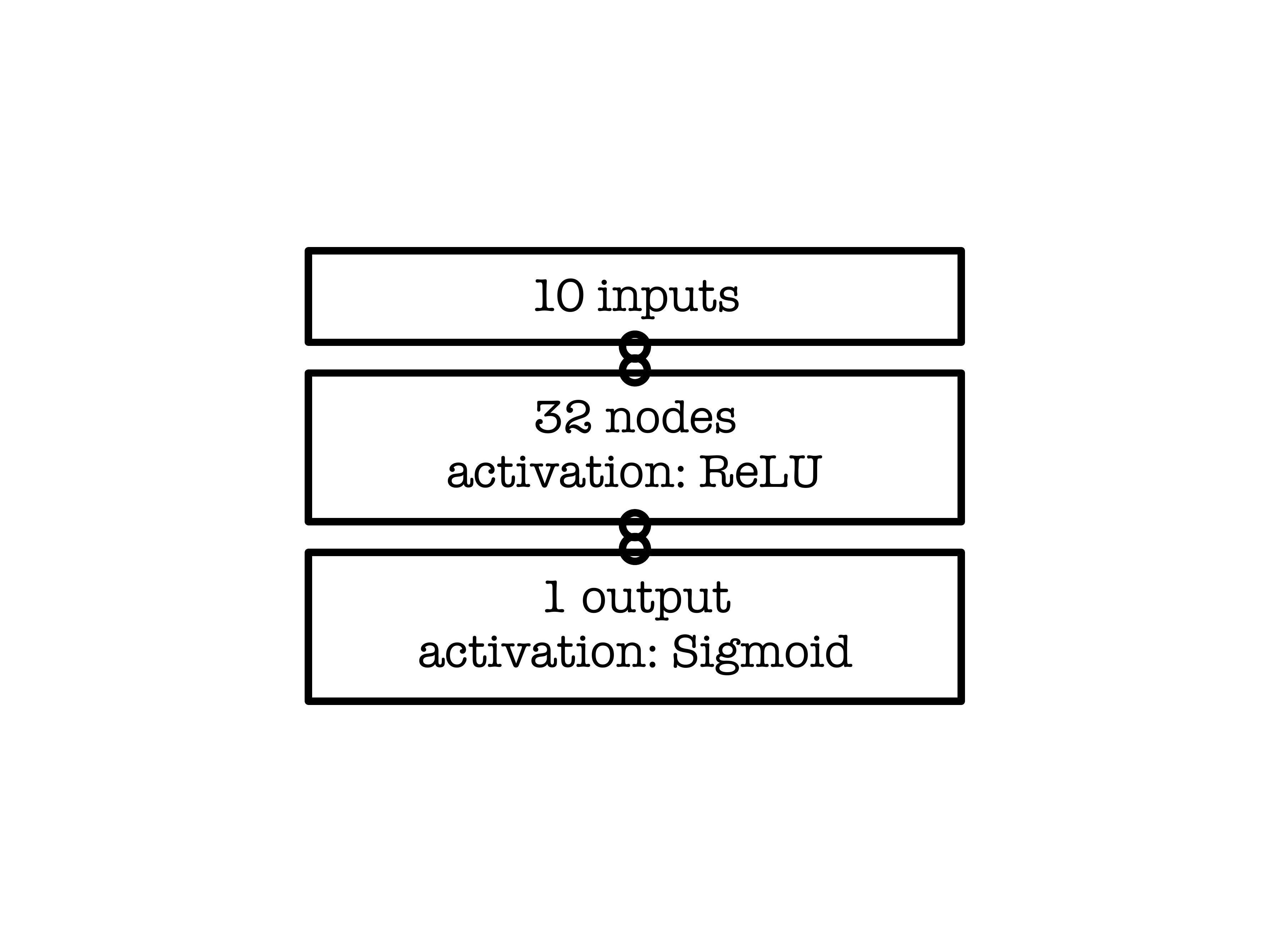}
\end{center}
\caption{Two neural network architectures for jet substructure classification. (Left) A three-hidden-layer model we use to categorize five classes of jets ($q$, $g$, $W$, $Z$, and $t$). (Right) A one-hidden-layer model used to identify top quarks, simplified for the FPGA implementation described in Sec.~\ref{sec:implementation}.}
\label{fig:arch}
\end{figure}

We train a neural network for the classification task of $q$, $g$, $W$, $Z$, and $t$ discrimination. The data are randomly split into training (60\%), validation (20\%), and testing (20\%) datasets. The input features are standardized by removing the mean and scaling to unit variance. The architecture, illustrated in Fig.~\ref{fig:arch} (left), is a fully-connected neural network with 
three hidden layers. The activation function for the hidden layers is ReLU~\cite{ReLU} while the output layer activation function is a softmax function to 
provide probabilities for each class. The categorical cross-entropy loss function is minimized with and without $L_1$ regularization of the weights (Sec.~\ref{sec:compression}) using the Adam algorithm~\cite{DBLP:journals/corr/KingmaB14} with an initial learning rate of $10^{-4}$ and a minibatch size of $1024$. The learning rate is halved if the validation loss fails to improve over 10 epochs.
Training is performed on an AWS EC2 P2 GPU instance~\cite{awsec2} with %the Deep Learning AMI~\cite{deeplearningami} and 
{\tt{Keras}}.
We also consider a simpler architecture with one hidden layer, see Fig.~\ref{fig:arch} (right), when studying the final FPGA implementation on a specific device.
This is described further in Sec.~\ref{sec:implementation}.

\begin{figure}[tbh!]
\begin{center}
\includegraphics[width=0.48\linewidth]{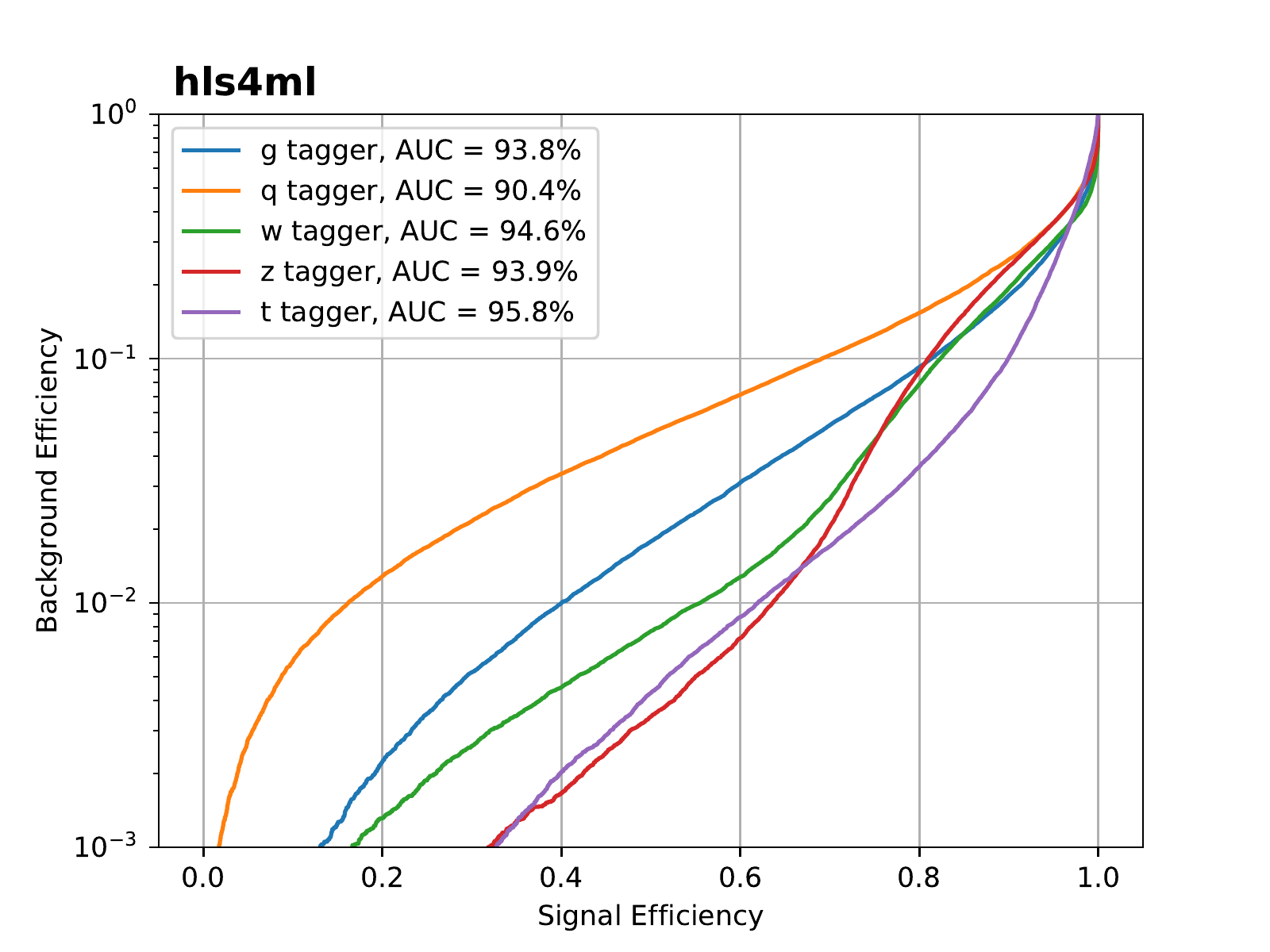}
\includegraphics[width=0.48\linewidth]{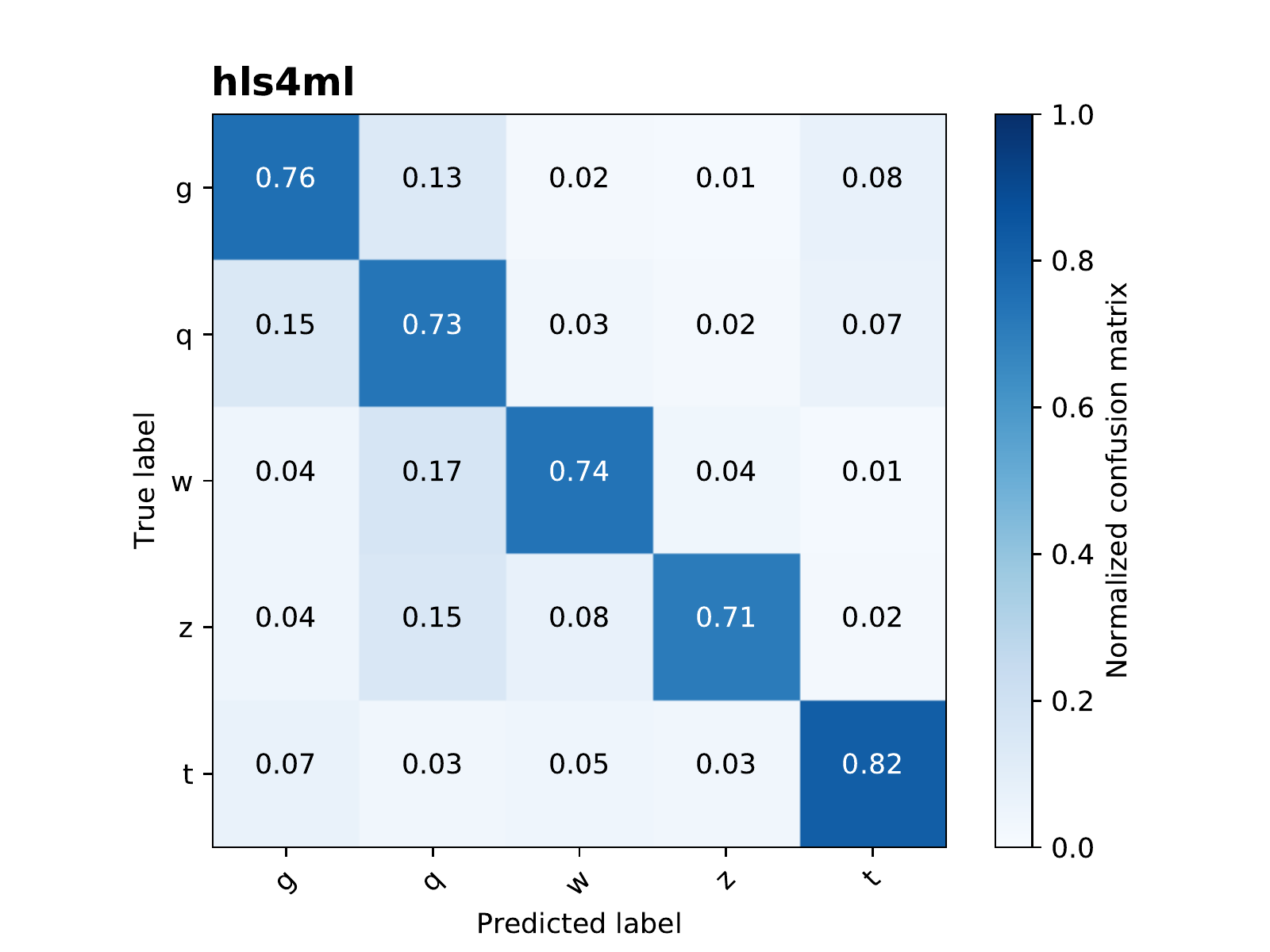}
\end{center}
\caption{Performance of the deep neural network classifier: (Left) signal efficiency versus mis-identification rate for quark, gluon, $W$ boson, $Z$ boson, and top quark jet identification.  The mis-identification rate is based on an equal admixture of the other non-signal jet types. (Right) The corresponding normalized confusion matrix for the five classes.}
\label{fig:roc1}
\end{figure}

The performance of the neural network classifier is shown in Fig.~\ref{fig:roc1}.
The general features of this performance plot are typical of jet substructure classification tasks.  
Top-quark jets, by virtue of their large mass and three-prong nature, have the best separation from the rest of the jet types.  
The $W$ and $Z$ jets are similar in performance because of their masses and two-prong nature while quark and gluon jets are notoriously challenging to classify~\cite{Asquith:2018igt}.
Given this multi-jet classifier performance, we explore how to implement such a neural network architecture in an FPGA using \hlsfml.

\subsection{Efficient network design}
\label{sec:hls4ml-design}
In Sec.~\ref{sec:hls4ml-concept}, we present a general description of the translation of a deep neural network into an FPGA implementation in Sec.~\ref{sec:hls4ml-concept}
and a specific network design for the task of jet substructure classification is introducted in Sec.~\ref{sec:hls4ml-substructure}. 
We now focus on tuning the network inference in a way that uses the FPGA resources efficiently and meets latency and pipelining constraints.

Neural network inference can be made efficient with the following techniques: compression, quantization, and parallelization. 
We summarize these ideas briefly:
\begin{itemize}
\item {\bf compression}: neural network synapses and neurons can be redundant; compression attempts to reduce the number of synapses or neurons thereby effectively reducing $N_{\rm multpliers}$ without suffering any performance loss;
\item {\bf quantization}: often 32-bit floating point calculations are not needed in the inference of a network to achieve optimal performance; quantization can reduce the precision of the calculations (weights, biases, etc.) in the neural network with no loss in performances; 
\item {\bf parallelization}: one can tune how much to parallelize the multiplications required for a given layer computation; in one extreme, all multiplications can be performed simultaneously using a maximal number of multipliers, while alternatively in the other extreme, one can use only one multiplier and perform the multiplications sequentially; between these extremes the user can optimize algorithm throughput versus resource usage.
\end{itemize}
In the following subsections, we describe in more detail the implementation and effect of these optimizations.

One important topic that we do not discuss is how the input features are computed before being fed to the neural network as this depends on the specific application.
For example, in the jet substructure case study we consider, the pre-computation of the expert features can be quite time consuming and resource intensive.
However, in other cases, the neural network may take raw detector inputs which require little preprocessing.
The consideration of the computation time of the inputs to the neural network is an important consideration in a more realistic scenario.
Additionally, it is important to consider the precision and range of the inputs; bit-shifting or translating the inputs to the appropriate range may be important to the performance of the algorithm.

\subsubsection*{Compression}
\label{sec:compression}

Network compression is a widespread technique to reduce the size, energy consumption, and overtraining of deep neural networks~\cite{DBLP:journals/corr/HanMD15}. Several approaches have been successfully deployed to compress networks, including~\cite{DBLP:journals/corr/abs-1710-09282}:
\begin{itemize}
\item {\bf parameter pruning}: selective removal of weights based on a particular ranking~\cite{NIPS1989_250,2017arXiv171201312L,DBLP:journals/corr/HanMD15}, 
\item {\bf low-rank factoriation}: using matrix/tensor decomposition to estimate informative parameters~\cite{6619199,NIPS2014_5544,DBLP:journals/corr/JaderbergVZ14,NIPS2013_5025,6638949}, 
\item {\bf transferred/compact convolutional filters}: special structural convolutional filters to save parameters~\cite{2016arXiv160207576C}, and
\item {\bf knowledge distillation}: training a compact network with distilled knowledge of a large network~\cite{Bucilua:2006:MC:1150402.1150464}.
\end{itemize}

Our approach is a simplified version of iterative parameter pruning and retraining~\cite{DBLP:journals/corr/HanPTD15,DBLP:journals/corr/HanMD15} with $L_1$ regularization, where the loss function $L$ is augmented with an additional penalty term,
\begin{align}
  L_\lambda(\vec w) &= L(\vec w) + \lambda \norm{\vec w}_1~.
\end{align}
$L_1$ regularization is known to produce sparse models, provide built-in feature selection~\cite{Ng:2004:FSL:1015330.1015435}, and is a readily available option in many machine learning workflows. In principle, training with $L_p$ regularization with $0\leq p<1$~\cite{2017arXiv171201312L} may improve the sparsity and performance of the model, but these regularizers are not always easy to implement.
While we take this simplified approach, we note that there are other, more sophisticated, approaches to compression in the literature which may yield even better results.

\begin{figure}[tbh!]
\centering
\includegraphics[width=0.96\linewidth]{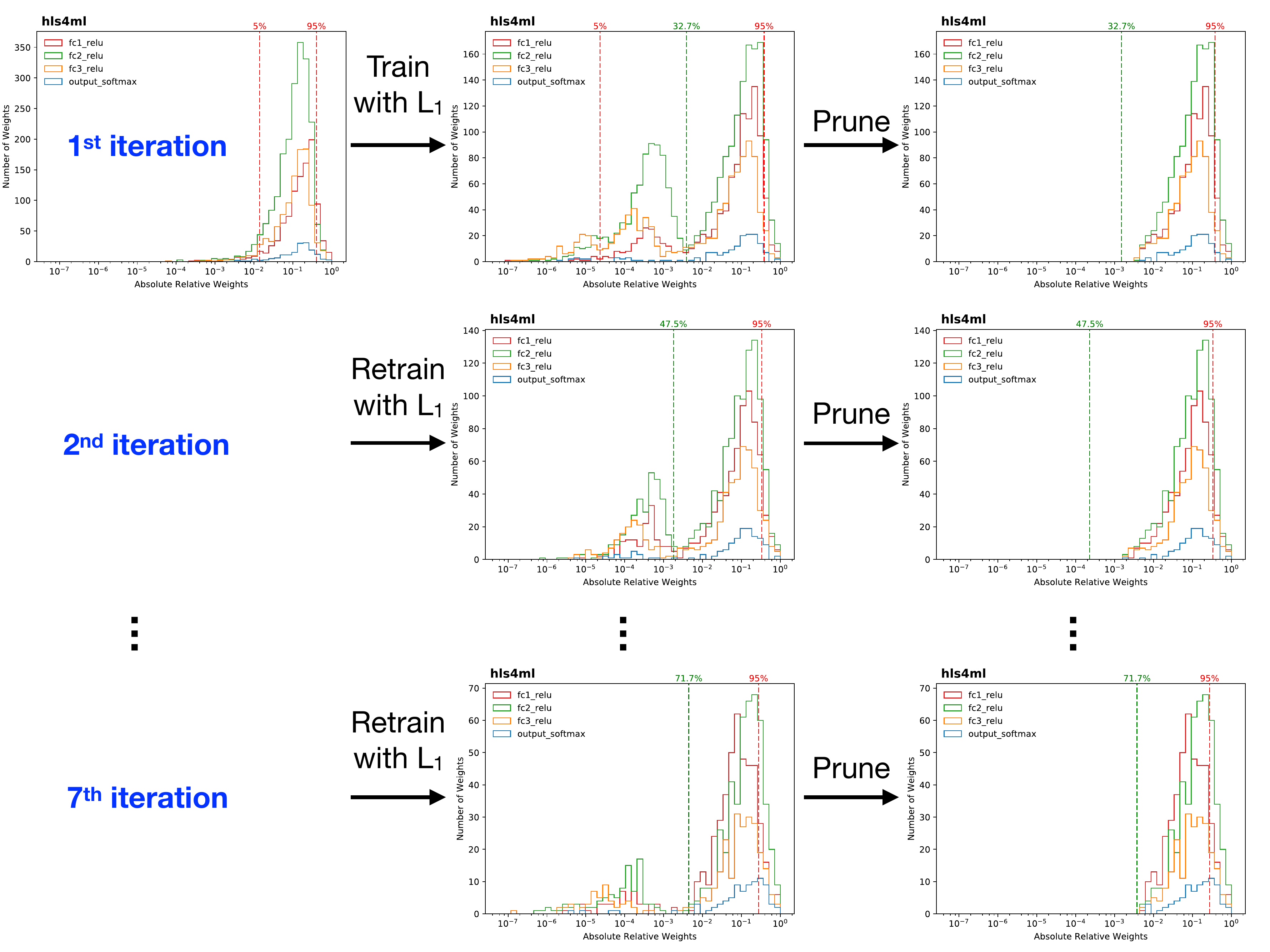}
\caption{Illustration of the iterative parameter pruning and retraining with $L_1$ regularization procedure. The distribution of the absolute value of the weights relative to the maximum absolute value of the weights is shown after each step of the pruning and retraining procedure. In the top left, the distribution before compression is shown, while in the bottom right, the distribution after compression is displayed.}
\label{fig:pruning}
\end{figure}

We train the model with $L_1$ regularization with $\lambda=10^{-4}$. We then
sort the weights based on their absolute value relative to the maximum
absolute value of the weights in a particular layer. 
With $L_1$ regularization we see two separate sub-populations of weights with one at smaller values and one at larger values. 
Weights falling below a certain percentile, corresponding to the smaller-value sub-population, are removed.
Next, we retrain the model again with $L_1$ regularization while constraining the previously pruned weights to remain zero. 
We stop after seven iterations of this procedure at which point the sum of the pruned weight sub-population is 3\% of the original summed weight population and the model is compressed by 70\% ($3051$ weights pruned out of $4389$ original weights and biases). Fig.~\ref{fig:pruning} illustrates this procedure. The top left of Fig.~\ref{fig:pruning} shows the distribution of the weights before compression. From the top left to the bottom right, the arrows indicate the following steps of the pruning and retraining procedure and the resulting distribution of weights is shown.
Finally, in the bottom right, we present the final distribution of the weights after compression. We observe no significant change in the pruned network performance when compared with the original. 

\subsubsection*{Quantization}
\label{sec:quantization}

Quantized~\cite{DBLP:journals/corr/GongLYB14,wu2016quantized,37631,DBLP:journals/corr/GuptaAGN15,DBLP:journals/corr/HanMD15} and even binarized~\cite{NIPS2015_5647,NIPS2016_6573,DBLP:journals/corr/RastegariORF16,DBLP:journals/corr/MerollaAAEM16} neural networks have been studied in detail as an additional way to compress neural networks by reducing the number of bits required to represent each weight.
FPGAs provide considerable freedom in the choice of data type and precision.  
Both are important to consider to prevent wasting FPGA resources and incurring additional latency.  
In \hlsfml \ we use fixed point arithmetic, which uses less resources and latency than floating point arithmetic.  

\begin{figure}[tbh!]
\centering
\includegraphics[width=0.48\linewidth]{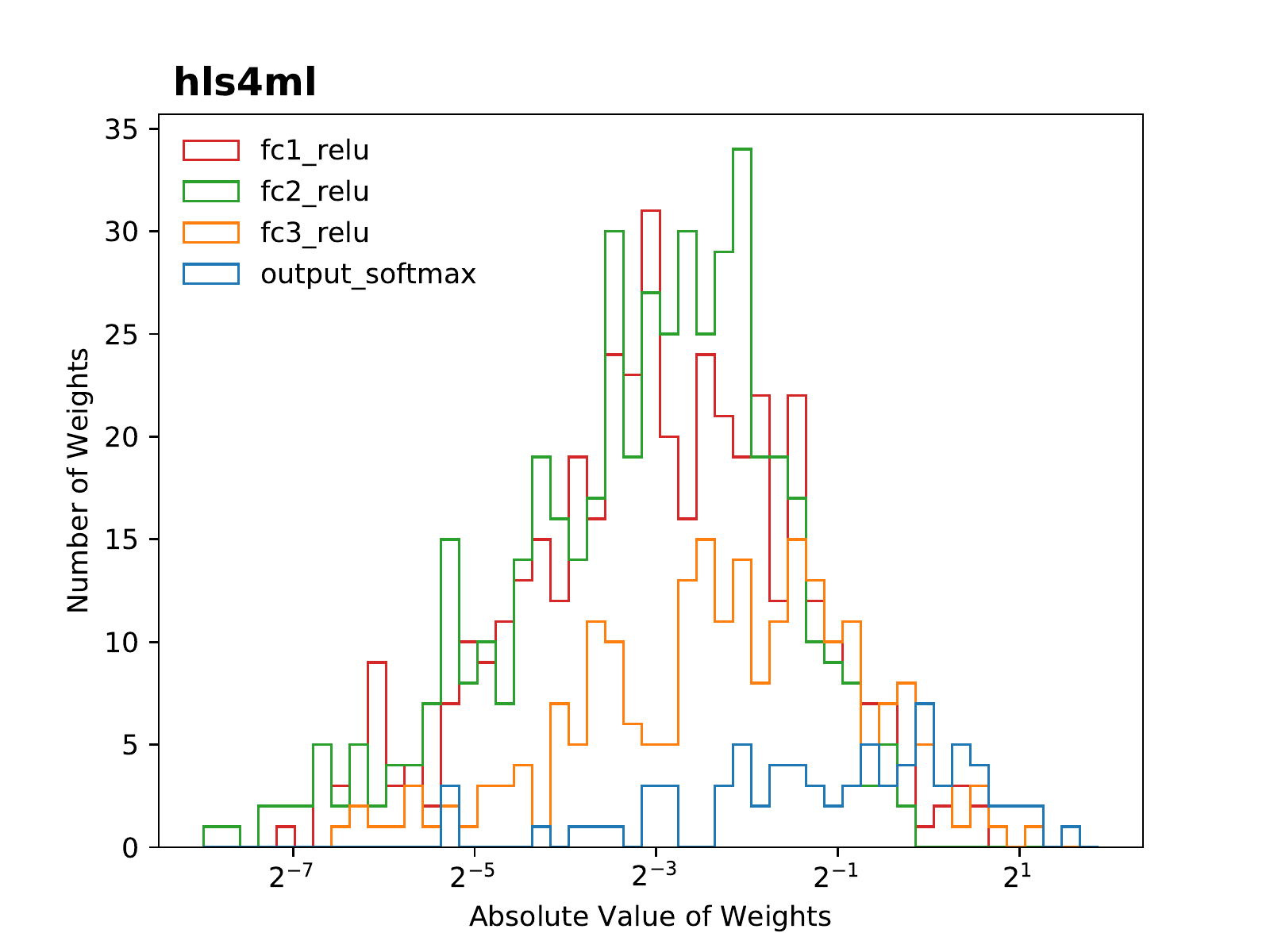}
\caption{Distribution of the absolute value of the weights after compression.}
\label{fig:weights}
\end{figure}

The inputs, weights, biases, sums, and outputs of each layer (see Eq.~\ref{eq:vcalc}) are all represented as fixed point numbers. 
For each, the number of bits above and below the binary point can be configured for the use case.  
It is broadly observed that precision can be reduced significantly without causing a loss in performance~\cite{DBLP:journals/corr/GuptaAGN15},
but this must be done with care.  
In Fig.~\ref{fig:weights}, we show the distribution of the absolute value of the weights after the compression described in Sec.~\ref{sec:compression}.
In this case, to avoid underflow/overflow in the weights, at least three bits should be assigned above the binary point --- two to envelope the largest absolute value and one for the sign.  
The neuron values, $\vec{x}_m$, and intermediate signals in the FPGA used to compute them may require more bits to avoid underflows/overflows.
We determine the number of bits to assign {\it below} the binary point by scanning physics performance as a function of the bit precision.

Reducing precision saves resources used for signal routing as well as resources and latency used for mathematical operations.
For many applications, the limiting FPGA resource will be the number of DSPs, which are used primarily for multiplications.
The number of DSPs used per multiplier depends on the precision of the numbers being multiplied and can change abruptly.  
For example, one Xilinx DSP48E1 block~\cite{dsp48e1} can multiply a 25-bit number with an 18-bit number, 
but two are required to multiply a 25-bit number with a 19-bit number.  
Similarly, the latency of multipliers increases with precision, though they can remain pipelined.
Detailed exploration of the effect of calculation precision is presented in Sec.~\ref{sec:results}. 

As mentioned in Sec.~\ref{sec:hls4ml-concept}, non-trivial activation functions 
are precomputed for a range of input values and stored in BRAMs.  
The binning within this range and the output bit width are configurable in \hlsfml. Lastly, we note that additional methods exist to further compress the network architecture through quantization that have not been explored in this paper~\cite{DBLP:journals/corr/GongLYB14,DBLP:journals/corr/RastegariORF16}. In particular, retraining the network with a quantized precision in the training can lead to equivalent performance with significantly smaller weight precision~\cite{DBLP:journals/corr/ZhuHMD16}. We leave investigations of these approaches for further work. 

\subsubsection*{Parallelization}
\begin{figure}[tbh!]
\centering
\includegraphics[width=0.8\linewidth]{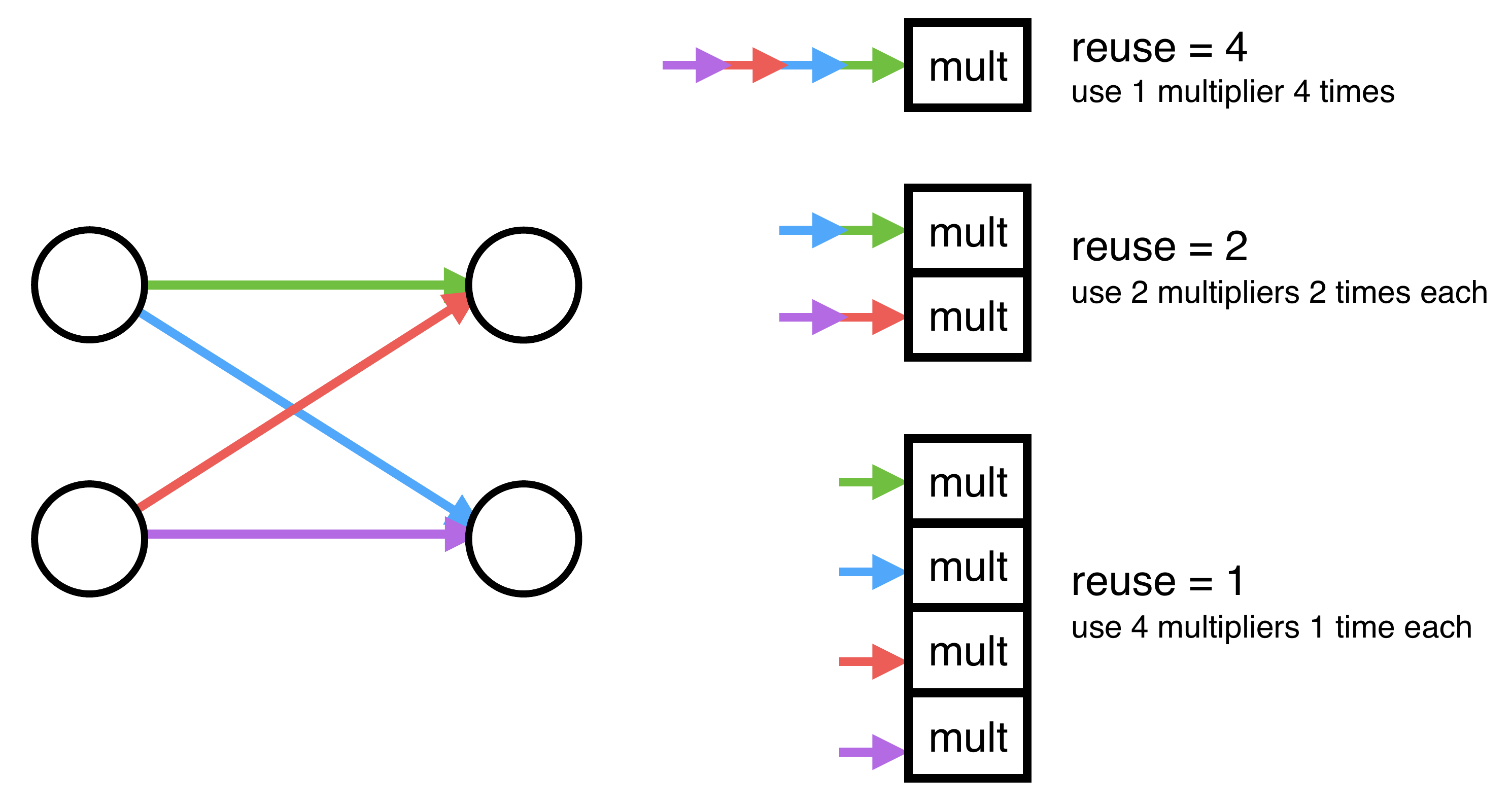}
\caption{Illustration of multiplier resource usage for different values of reuse factor. The left drawing shows the case of two neuron pairs being connected by 4 connections, implying four multiplications to be performed. The right drawing shows how to perform these multiplications, from fully serial (top) to fully parallelized (bottom).}
\label{fig:reuse}
\end{figure}

The trade-off between latency, throughput and FPGA resource usage is determined by the parallelization of the inference calculation.
In \hlsfml, this is configured with a multiplier ``reuse factor'' that sets the number of times a multiplier is used 
in the computation of a layer's neuron values.  
With a reuse factor of one, the computation is fully parallel.  
With a reuse factor of $R$, $1/R$ of the computation is done at a time with a factor of $1/R$ fewer multipliers. 
This is illustrated in Fig.~\ref{fig:reuse}.

FPGA multpliers are pipelined; therefore, the latency of one layer computation, $L_{m}$, is approximately
\begin{align}
L_{m} &= L_{\rm mult} + (R-1) \times {\rm II}_{\rm mult} + L_{\rm activ}~,
\label{eq:latency_one_layer}
\end{align}
where $L_{\rm mult}$ is the latency of the multiplier, ${\rm II}_{\rm mult}$ is the initiation interval of the multiplier, 
and $L_{\rm activ}$ is the latency of the activation function computation.
Equation~\ref{eq:latency_one_layer} is approximate because, in some cases, additional latency can be incurred for signal routing,
for instance in the addition of multiplication results contributing to a neuron value. 

As discussed in Sec.~\ref{sec:hls4ml-concept}, we implement each layer calculation independently and sequentially.
The calculation of one layer cannot be initiated until the calculation of the previous layer has completed.  
Therefore, the total latency is equal to the sum of latencies of each layer plus the latency required to connect the layers.
The number of inferences completed per unit time is inversely proportional to the reuse factor.

% %%%%%%%%%%%%%%%%%%%%%%%%%%%%%%%%%%%%%%%%%%%%%%%%%%%%%%%%%%%%%%%%%%%%%%%%%%%%%%%%%%%%%%%%%%%%%%%%
% % R E S U L T S
% %%%%%%%%%%%%%%%%%%%%%%%%%%%%%%%%%%%%%%%%%%%%%%%%%%%%%%%%%%%%%%%%%%%%%%%%%%%%%%%%%%%%%%%%%%%%%%%%
%\clearpage
\section{Performance and implementation}
\label{sec:results}
%%%%%%%%%%%%%%%%%%%%%%%%%%%%%%%%%%%%%%%%%%%%%%%%%%%%%%%%%%%%%%%%%%%%%%%%%%%%%%%%%%%%%%%%%%%%%%%%
In this section, we quantify the results from the HLS translation and
optimization of the jet substructure neural network described in Sec.~\ref{ss:full3layerNN} as a function of
the three basic principles described in the previous section: compression, quantization, and parallelization.
First we discuss the classification performance of the neural network when implemented in firmware in Sec.~\ref{sec:perf}.
Then in Sec.~\ref{sec:resources}, we quantify the HLS synthesis in terms of FPGA resource usage and latency.  
The combination of these two metrics, classification and firmware performance, define how to optimally implement neural networks into FPGA hardware for a given application.
Finally, in Sec.~\ref{sec:implementation}, we discuss the
implementation for a specific FPGA and compare the actual resource usage to
the estimates from Vivado HLS, which can be obtained much more quickly.

\subsection{Classification performance}
\label{sec:perf}
\begin{figure}[tbh!]
\begin{center}
\includegraphics[width=0.48\linewidth]{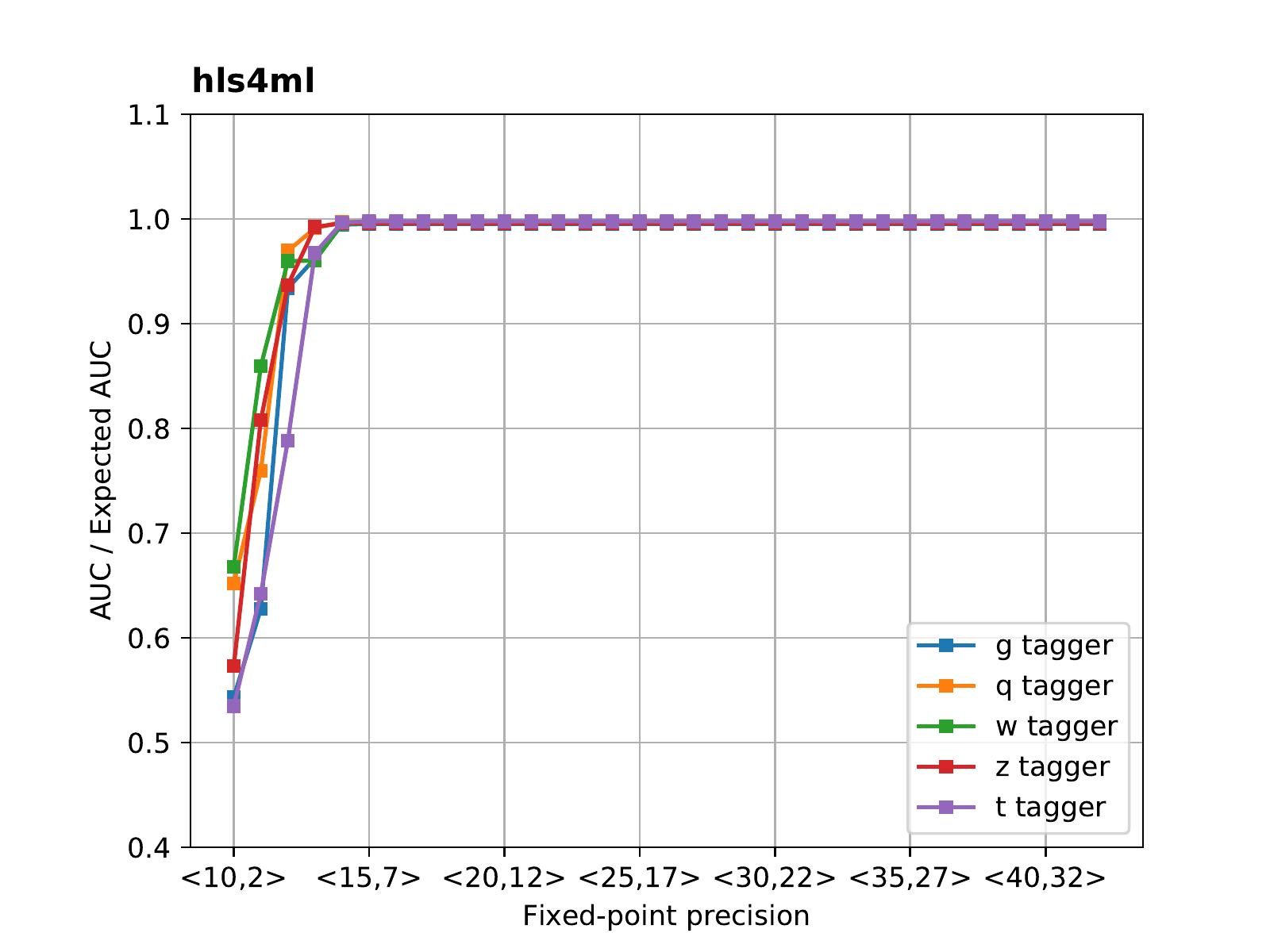}
\includegraphics[width=0.48\linewidth]{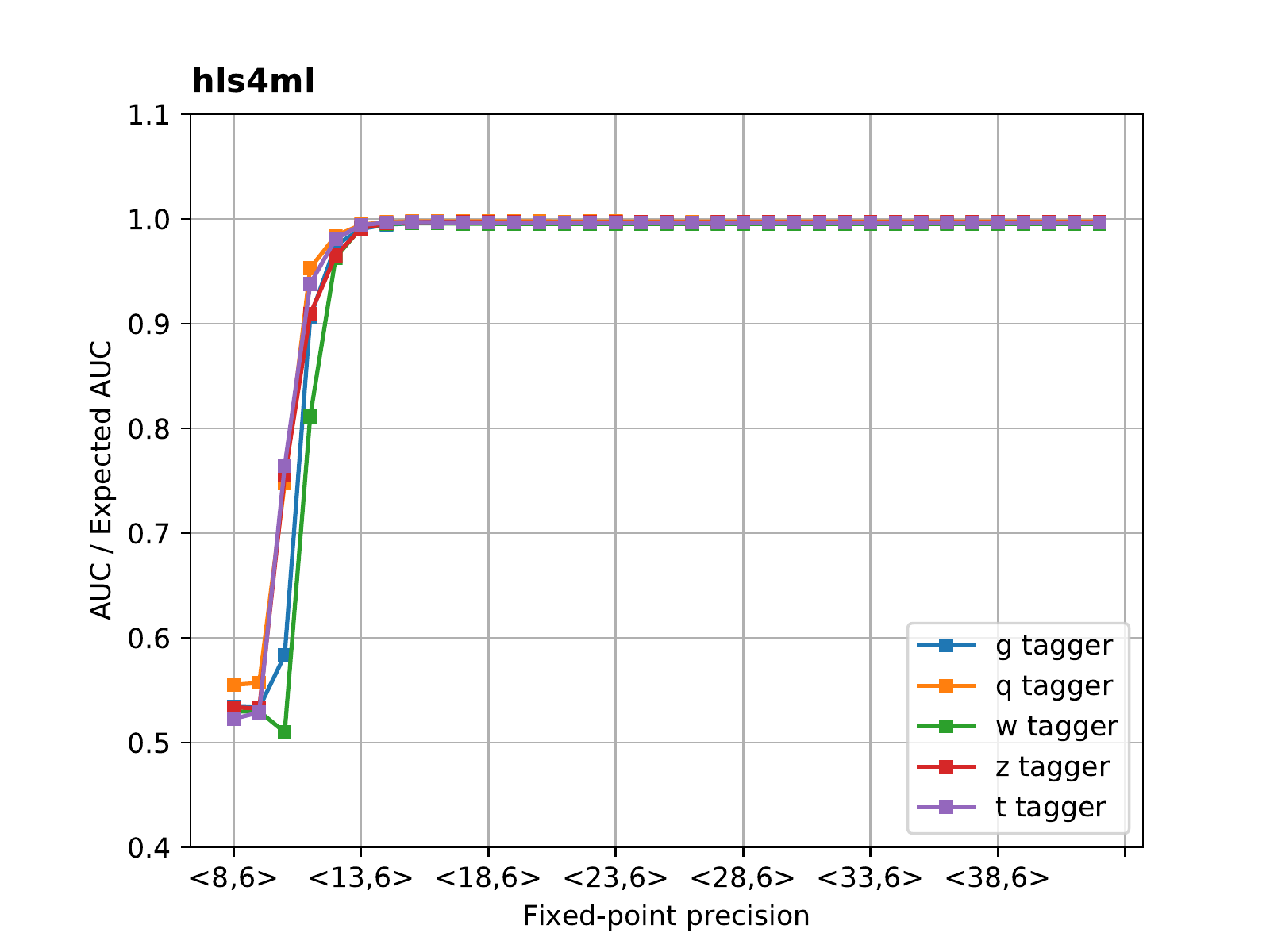}
\end{center}
\caption{ Ratios of the fixed point AUC and Expected AUC versus fixed point precision for the fully connected three-hidden-layer network. 
Optimal performance with no loss of classification power corresponds to ratios of 1.
(left) The number of integer bits is scanned.
(right) The number of integer bits is fixed to six, and the number of fractional bits is scanned.
The various colored lines are AUC performance for
different jet substructure taggers ($q$,$g$,$W$,$Z$,$t$).}
\label{fig:auc-perf}
\end{figure}

To quantify the performance of our five-output classifier, we
use the $\mathrm{AUC}$ metric, or area under the Receiver Operating
Characteristic (ROC) curve.
The ROC curve is given by the background rejection versus signal efficiency computed from sequential cuts on the classifier output,
where background rejection is $(1-\mathrm{background~efficiency)}$, as illustrated in Fig.~\ref{fig:roc1}. 
We denote the $\mathrm{AUC}$ achieved by a full 32-bit floating point inference of the neural network as $\mathrm{Expected~AUC}$.
We evaluate the neural network with fixed point precision denoted by {\tt <X,Y>} where {\tt Y} is 
the number of bits representing the signed number above the binary point (i.e. the integer part), and {\tt X} is the total number of bits.
We perform two scans -- one where we fix the number of integer bits and one where we fix the number fractional bits. 
The results are illustrated in Fig.~\ref{fig:auc-perf} where the scan of the integer bits is on the left
and the scan of the fractional bits is on the right.

Optimal performance with no loss of classification power corresponds to
$\mathrm{AUC}/\mathrm{Expected~AUC}~=~1$.  Fig.~\ref{fig:auc-perf} shows that with 
fixed point calculations and a sufficient number of bits, the Expected AUCs can be reproduced with negligible loss in performance. 
The number of integer bits is chosen to be just above the point where underflows/overflows do not occur and $\mathrm{AUC}/\mathrm{Expected~AUC} = 1$.
With this number of integer bits, we then scan in the number of fractional bits. % -- as is shown in Fig.~\ref{fig:auc-perf} (right). 
Optimal performance is achieved with about 16 bits in total.  

\begin{figure}[tbh!]
\begin{center}
\includegraphics[width=0.60\linewidth]{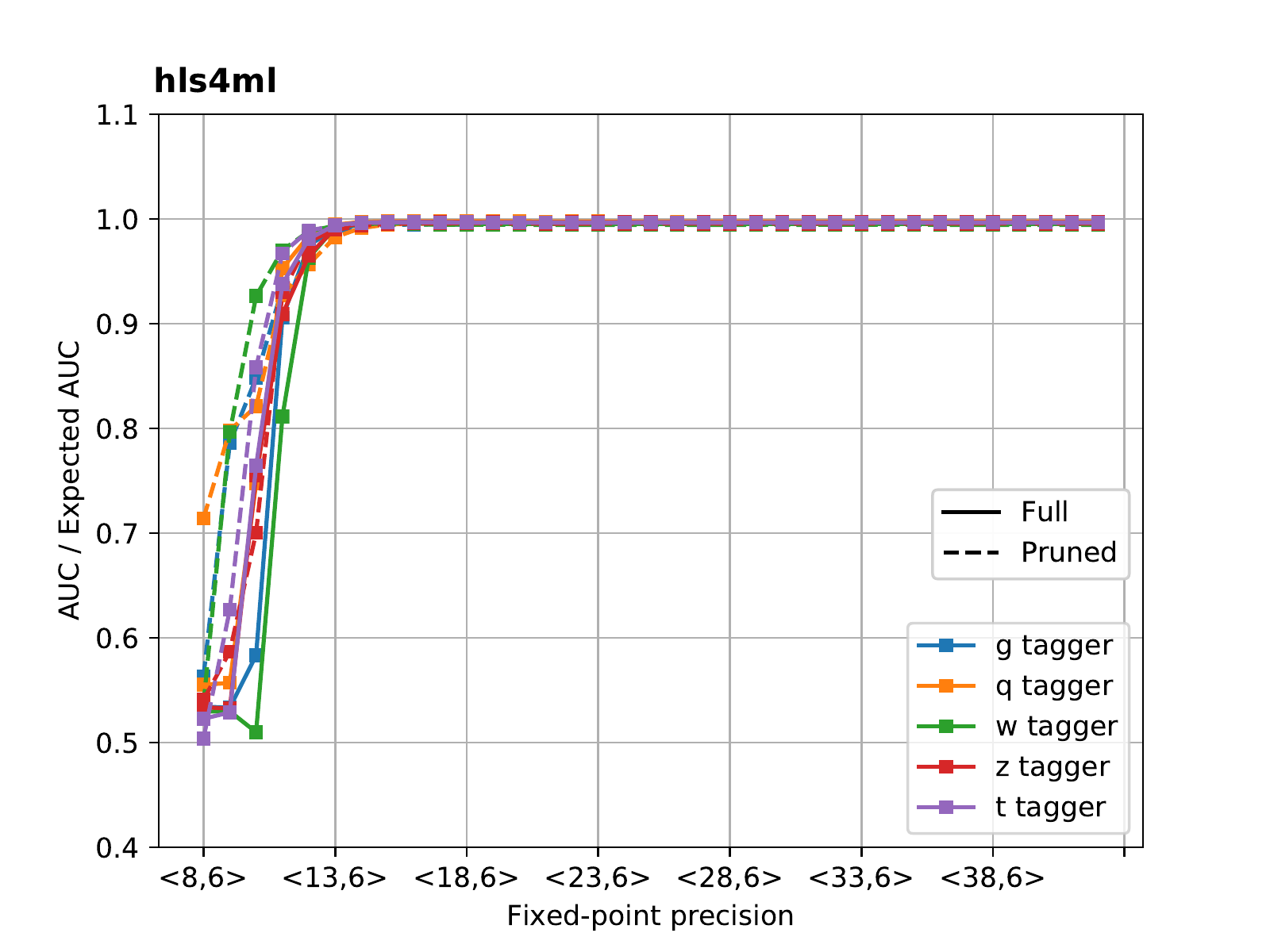}
\end{center}
\caption{ Ratios of the fixed point AUC and expected AUC versus fixed point precision. 
The solid lines represent the fully connected network, while the dash lines denote the compressed network.}
\label{fig:auc-prun}
\end{figure}

We perform similar scans to compare the compressed three-hidden-layer model AUC with that of the uncompressed model.  
Agreement with the Expected AUC occurs at roughly the same precision, as shown in Fig.~\ref{fig:auc-prun}. 

\subsection{Latency and resource estimates in HLS}
\label{sec:resources}
We now explore how the FPGA resources required by the model are influenced by
\begin{itemize}
\item {\bf compression}, the three-hidden-layer model with 70\% of the parameters pruned;
\item {\bf quantization}, the precision of the inputs, weights, and biases; for this particular network we focus on scans of fixed point precision {\tt <X,6>} based on our discussion in Sec.~\ref{sec:perf}.  We scan above the point where we reach optimal performance to show the benefits of quantization and the resource usage one would expect when higher precision is required.
\item {\bf parallelization}, the number of times a given multiplier is used for a layer computation; using a multiplier once is the most parallel (and quickly) a layer can be computed and is what we call a {\it reuse factor} of 1. 
\end{itemize}
With these variables as handles on how to control the implementation of the network, %we study the firmware implementation.
we monitor the following firmware implementation metrics:
\begin{itemize}
\item {\bf resources}: DSPs, FFs, and LUTs;
\item {\bf latency}: the time it takes to compute the full network;
\item {\bf initiation interval}: the time before a new set of inputs can be accepted.
\end{itemize}
At the moment we do not probe the block RAM (BRAM) usage, which is only used to store precomputed activation function values.   
Storing and accessing neural network weights from BRAMs, for instance, leads to latencies longer than the requirements for the first stages of LHC L1 triggers.
For longer latency tasks %and pipeline applications
, e.g. HLT applications, the capabilities of \hlsfml~can be expanded to allow for weight storage in BRAMs.

The results presented below are synthesized for a Xilinx Kintex Ultrascale FPGA with part number xcku115-flvb2104-2-i.
The clock frequency is fixed at 200~MHz, which is typical for the first stages of LHC triggers.
There can be variations in results versus clock frequency, but in the $\mathcal{O}$(100~MHz) range, we find variations are negligible.
Resource usage estimates are taken from the Vivado HLS synthesis step
and are found to be conservative in general when compared to
implementation as we will discuss in Sec.~\ref{sec:implementation}. % Citation?
While conservative, the short time required to make HLS resource estimates makes them useful for
rapidly prototyping different network designs and deriving useful trends.
Discussion of our results' dependence on the version of Vivado HLS,
the specific FPGA, and the final implementation in the FPGA are
discussed in Sec.~\ref{sec:implementation}.

%%%%%%%%%%%%%%%%%%%%%%%%%%%%%%%%%%%%%%%%%%%%%%%%%%%%%%%%%%%%%%%%%%%%%%%%%%%%%%%%%%%%%%%%%%%%%%%%%%%%%%
\subsubsection*{Resources with compression}

\begin{figure}[tbh!]
\begin{center}
\includegraphics[width=0.48\linewidth]{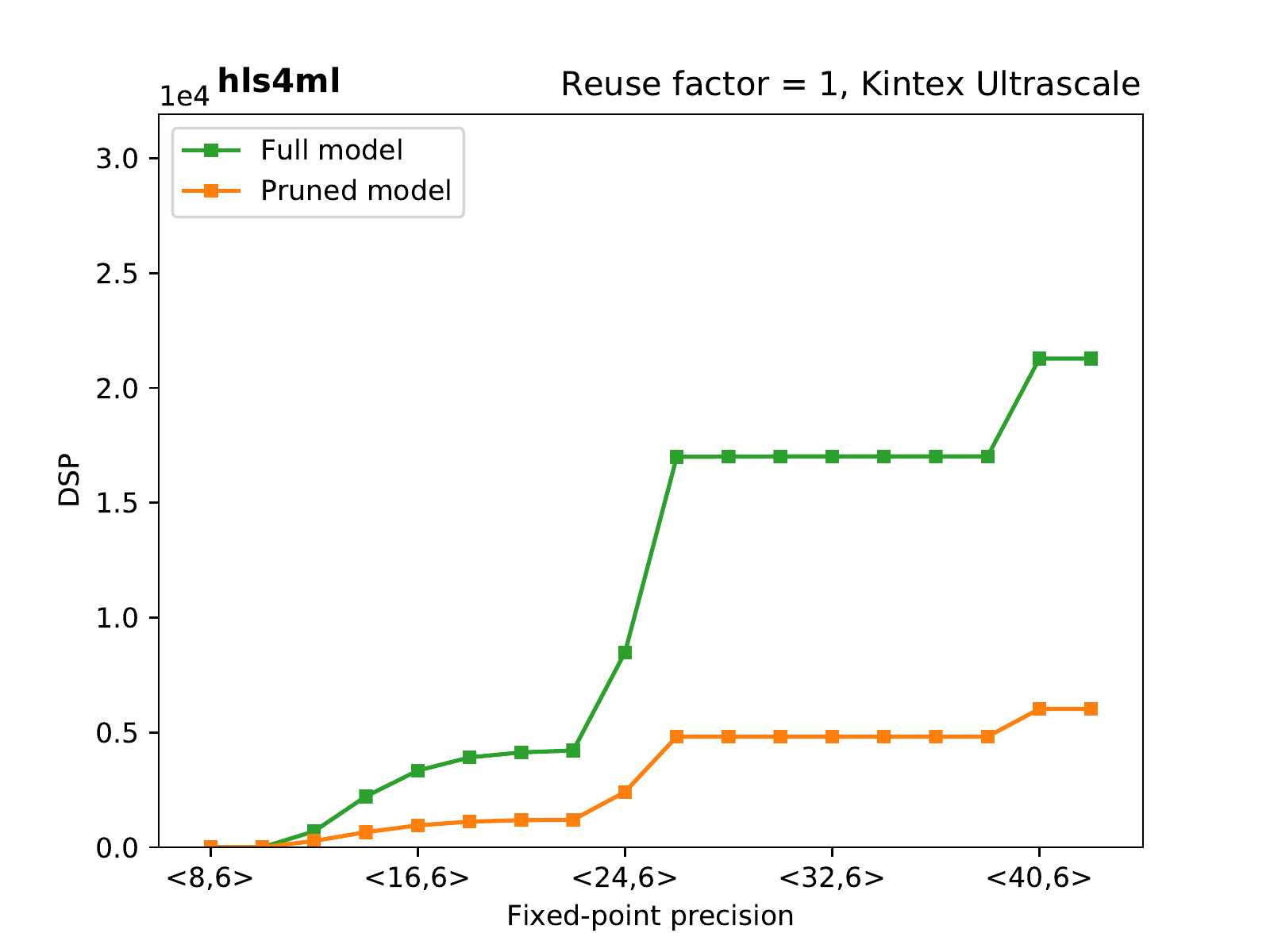}
\includegraphics[width=0.48\linewidth]{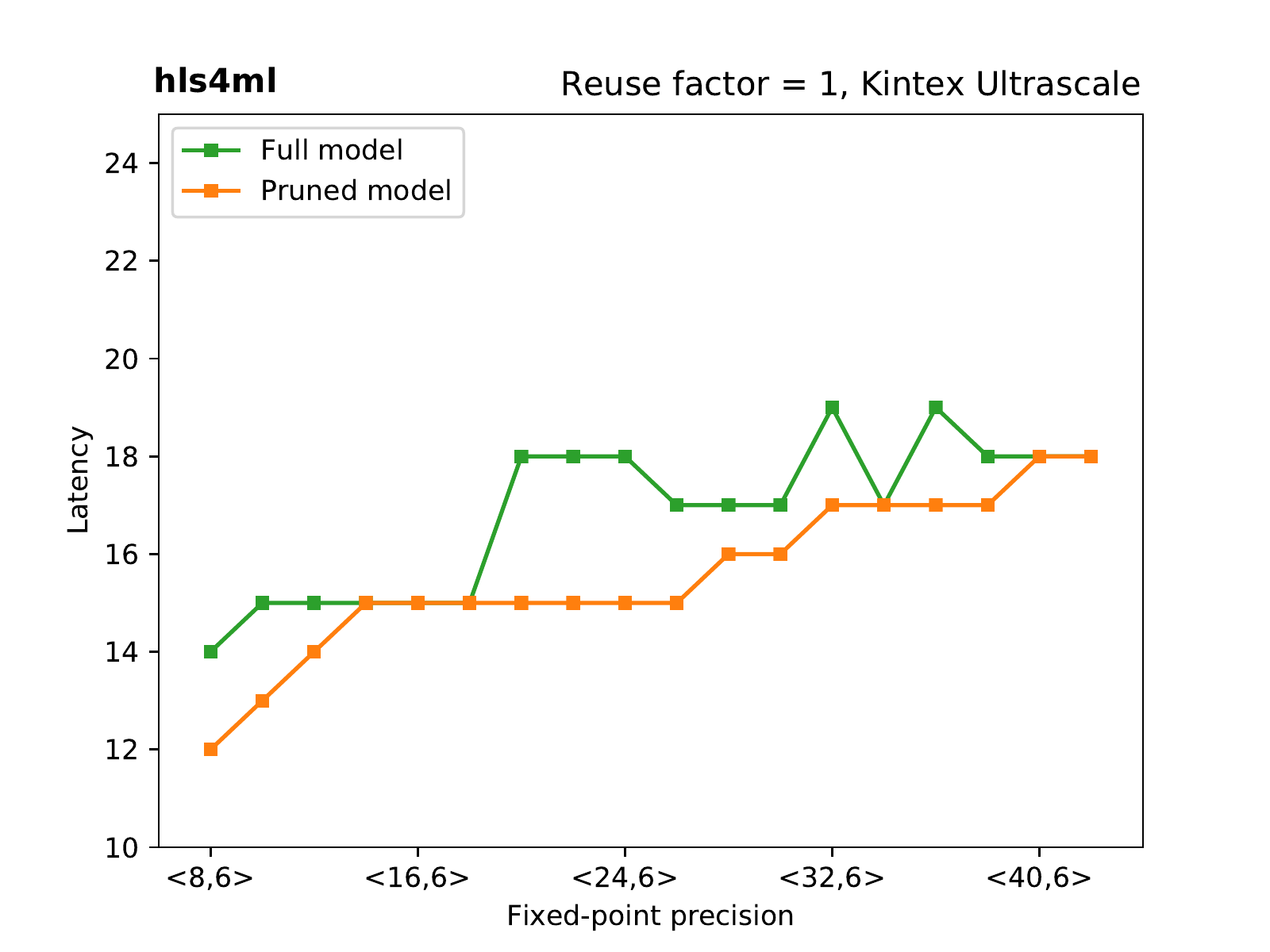}
\end{center}
\caption{A comparison between the compressed and uncompressed models, with a reuse factor of 1 for DSP usage (left) and latency in clock cycles for a 200~MHz clock frequency (right).  The x-axis is a scan in the fixed-point precision of the model and demonstrates how resource usage changes as a function of the precision of the calculations in the network inference.}
\label{fig:perf-pruned}
\end{figure}

We first explore the effect of compression on the FPGA resources required by the neural network.
Because the compression is typically part of the training workflow, we consider it separately from the other optimization handles.
Looking at the DSP usage and the algorithm latency in Fig.~\ref{fig:perf-pruned}, we show the
difference between our compressed and uncompressed neural network models.  
In both cases, we consider the network maximally parallelized ({\it reuse factor} of 1).
With the weights stored in programmable logic, sparse matrix multiplication is handled trivially and zero-weight multiplications are optimized out of the network FGPA implementation.   We find this
to be a very attractive feature of HLS though more sophisticated
compression techniques like those described
in~\cite{2016arXiv160201528H} may require more study.

As shown in Fig.~\ref{fig:perf-pruned} (left), the DSP usage is drastically
reduced for the compressed model compared to the
original network by an amount that is proportional to the 70\% compression rate described in Sec.~\ref{sec:hls4ml-design}.
In addition, the DSP usage increases as the fixed-point precision increases.
The increases are not smoothly varying because they depend on the DSP design precisions.
On the right of Fig.~\ref{fig:perf-pruned}, we present the latency of the algorithm in clock cycles for a 200~MHz clock frequency.
Because the network still has the same structure, in terms of the number of hidden layers, the latency is approximately the same in the compressed and uncompressed models.
Note that the total latency to infer the model is approximately 15
clock cycles which translates to 75~ns, well within the latency budgets of the first stages of LHC triggers.

To summarize the results of the HLS synthesis of the compressed and
uncompressed models, we report some vital statistics in Table~\ref{tab:comp-compress}.
We note the reduced resources while maintaining the same performance,
latency, and initiation interval.
\begin{table}[tbh!]
\centering
\begin{tabular}{|c|c|c|}	\hline
\textbf{Network} & Uncompressed network & Compressed network \\
\hline \hline
AUC~/~Expected AUC & 99.68\% & 99.55\% \\
\hline
Parameters & 4389 & 1338 \\
\hline
Compression factor & - & 3.3$\times$ \\
\hline
DSP48E & 3329 & 954 \\
\hline
Logic (LUT + FF) & 263,234 & 88,797 \\
\hline
Latency & 75~ns & 75~ns \\
\hline
\end{tabular}
\caption{
A summary of the vital statistics and HLS resource estimates of the uncompressed and compressed jet substructure tagging model with a network precision of fixed-point {\tt<16,6>} and fully pipelined with clock frequency of 200~MHz synthesized on a Xilinx Kintex Ultrascale FPGA.
}
\label{tab:comp-compress}
\end{table}

%%%%%%%%%%%%%%%%%%%%%%%%%%%%%%%%%%%%%%%%%%%%%%%%%%%%%%%%%%%%%%%%%%%%%%%%%%%%%%%%%%%%%%%%%%%%%%%%%%%%%%
\subsubsection*{Compressed three-hidden-layer Model Results}

\begin{figure}[tbh!]
\begin{center}
\includegraphics[width=0.60\linewidth]{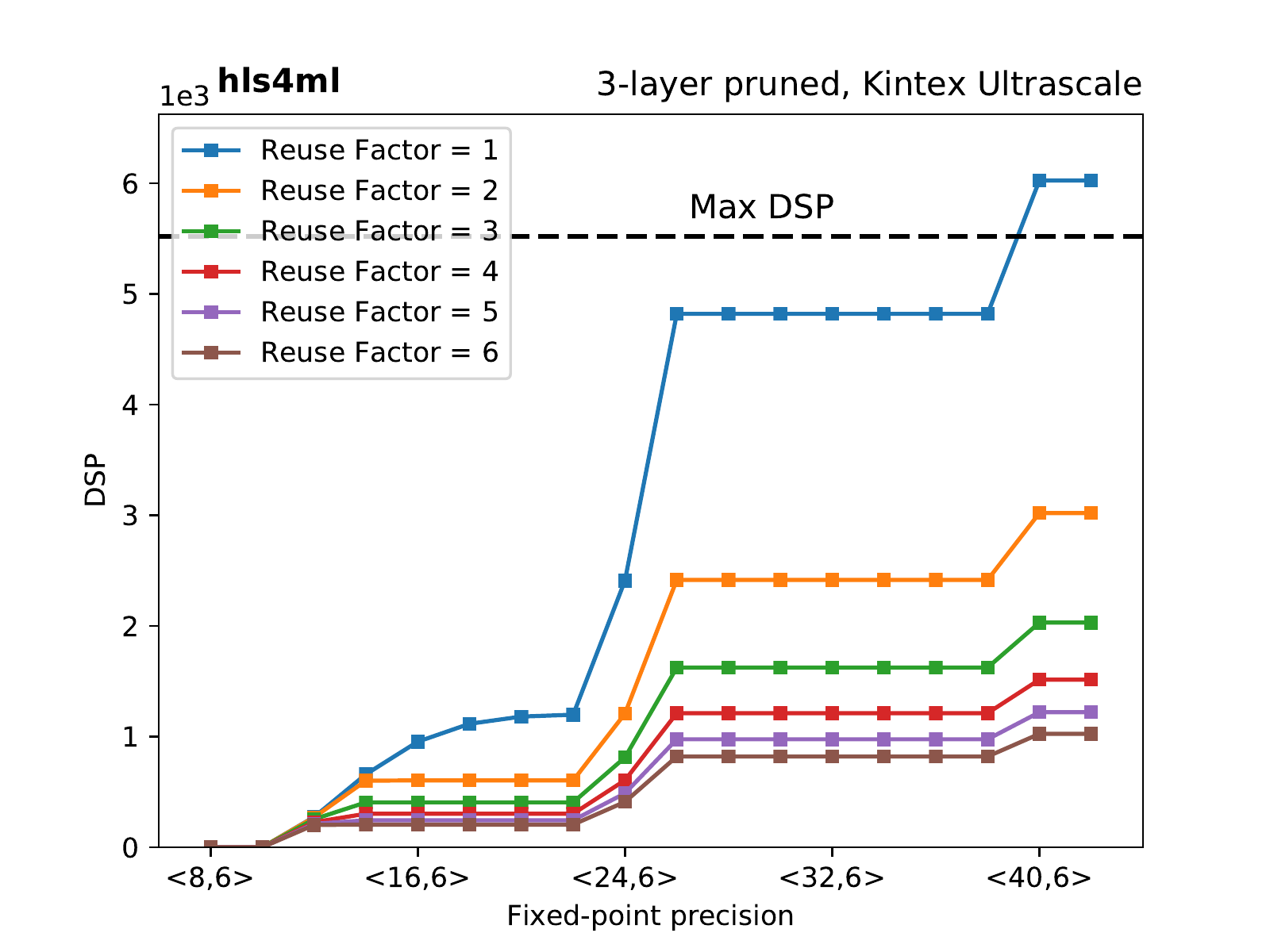}
\end{center}
\caption{DSP usage in the compressed three-hidden-layer model as a function of the network precision.  The various curves illustrate resource usage for different resource usage factors.}
\label{fig:perf-pruned-resources}
\end{figure}

\begin{figure}[tbh!]
\begin{center}
\includegraphics[width=0.48\linewidth]{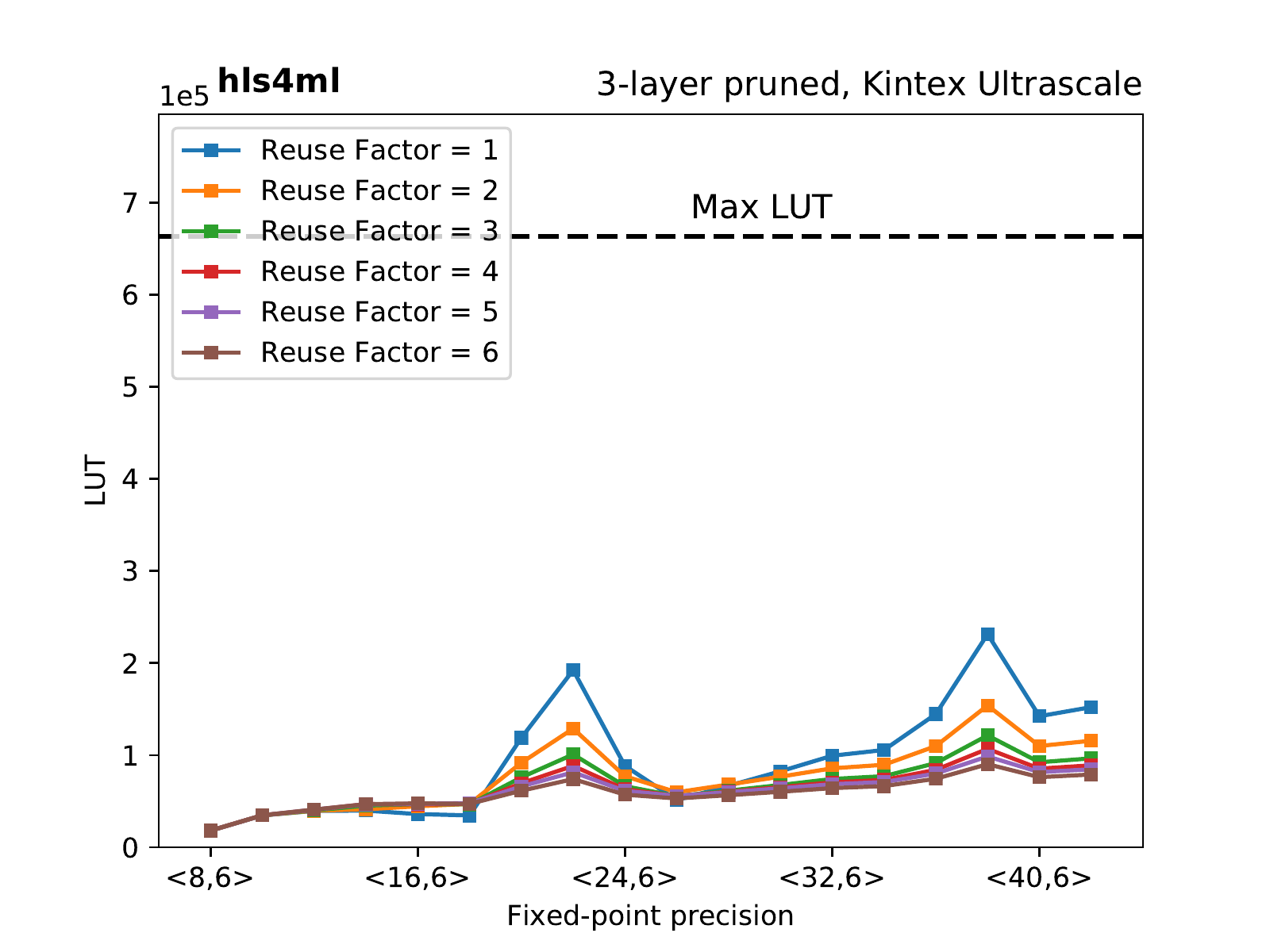}%
\includegraphics[width=0.48\linewidth]{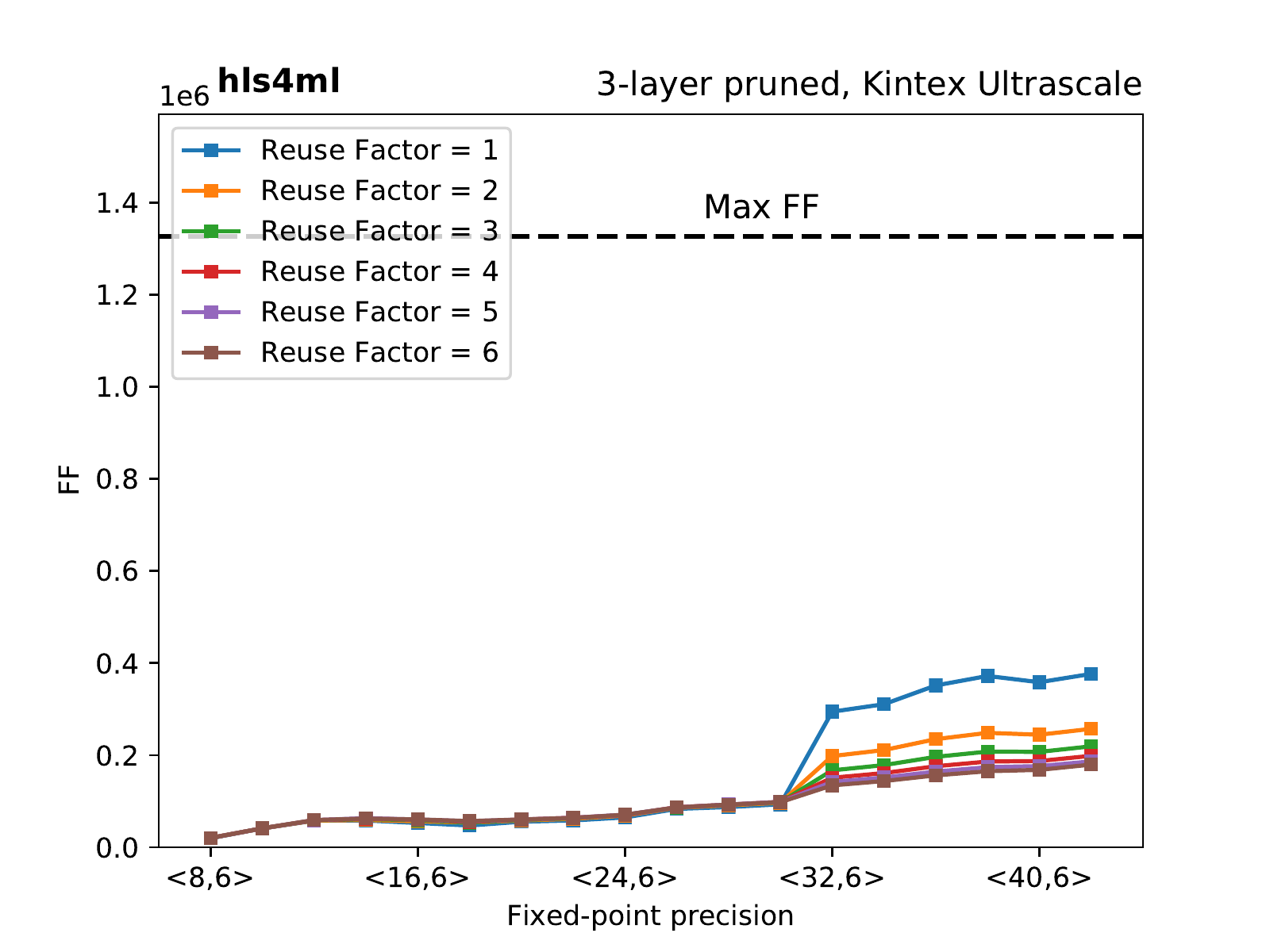}
\end{center}
\caption{LUT and FF usage in the compressed three-hidden-layer model as a function of the network precision.  The various curves illustrate resource usage for different resource usage factors.}
\label{fig:perf-pruned-resources-2}
\end{figure}

We now consider our compressed three-hidden-layer neural network model as the benchmark model for our use case and perform detailed scans of FPGA resources versus network precision and reuse factor.
In Fig.~\ref{fig:perf-pruned-resources} and Fig.~\ref{fig:perf-pruned-resources-2}, we examine the DSP, FF, and LUT usage as a function of precision of the fixed point calculations, {\tt <X,6>}.  
From the findings in Sec.~\ref{sec:perf}, we scan the number of fractional bits by scanning {\tt X} while fixing the integer bits at $Y=6$, guaranteeing no underflows/overflows.
Different curves are shown for different values of reuse factor. % which, as a reminder, defines how many times one reuses DSP multipliers in a given layer computation.

%In Fig.~\ref{fig:perf-pruned-resources}, we observe that the reuse factor controls the number of times a multiplier is used in the neural network.  
In Fig.~\ref{fig:perf-pruned-resources}, we show how the reuse factor is used to control the number of times a multiplier is used in the neural network.  
As the reuse factor increases, we are able to control the DSP usage proportionally to the reuse factor.
% The DSP usage ({\tt reuse = 3}) scales as the DSP usage ({\tt reuse = 1}) divided by 3.  
The DSP resource usage has steps in the resource usage as a function of the network precision.
This is consistent for all values of {\tt reuse} and comes from the precision of the Xilinx FPGA DSPs.
In the figure, we also indicate the maximum number of DSPs available in this particular Xilinx Kintex Ultrascale FPGA. 
In Fig.~\ref{fig:perf-pruned-resources-2}, the LUT (left) and FF (right) usage is shown.
For both the LUTs and the FFs, the resource usage relative to the
FGPA's capacity is small compared to that of the DSPs.
Additionally, we observe spikes in FF usage at the DSP precision limits. We find that they are removed when performing the implementation (discussed in Sec.~\ref{sec:implementation}).
We note a general trend across Fig.~\ref{fig:perf-pruned-resources} and Fig.~\ref{fig:perf-pruned-resources-2} that the {\tt reuse = 1} case can tend to deviate from the trends for the other cases when {\tt reuse > 1}.  We believe in the {\tt reuse = 1} case, HLS is able to do further optimizations on a single multiplier for a given network design, and it does not have that optimization freedom when a multiplier is tasked with multiple operations when {\tt reuse > 1}.

\begin{figure}[tbh!]
\begin{center}
\includegraphics[width=0.48\linewidth]{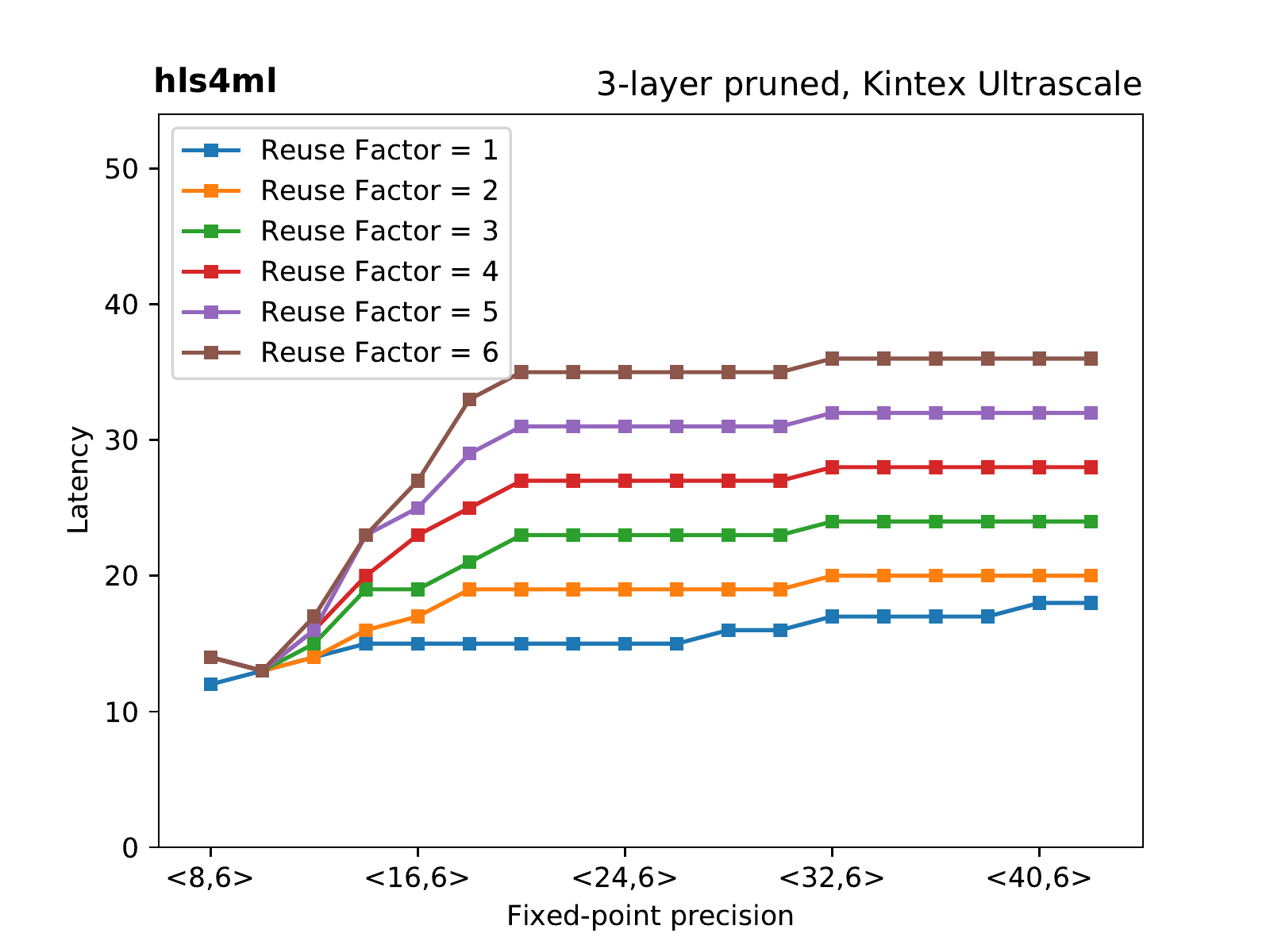}%
\includegraphics[width=0.48\linewidth]{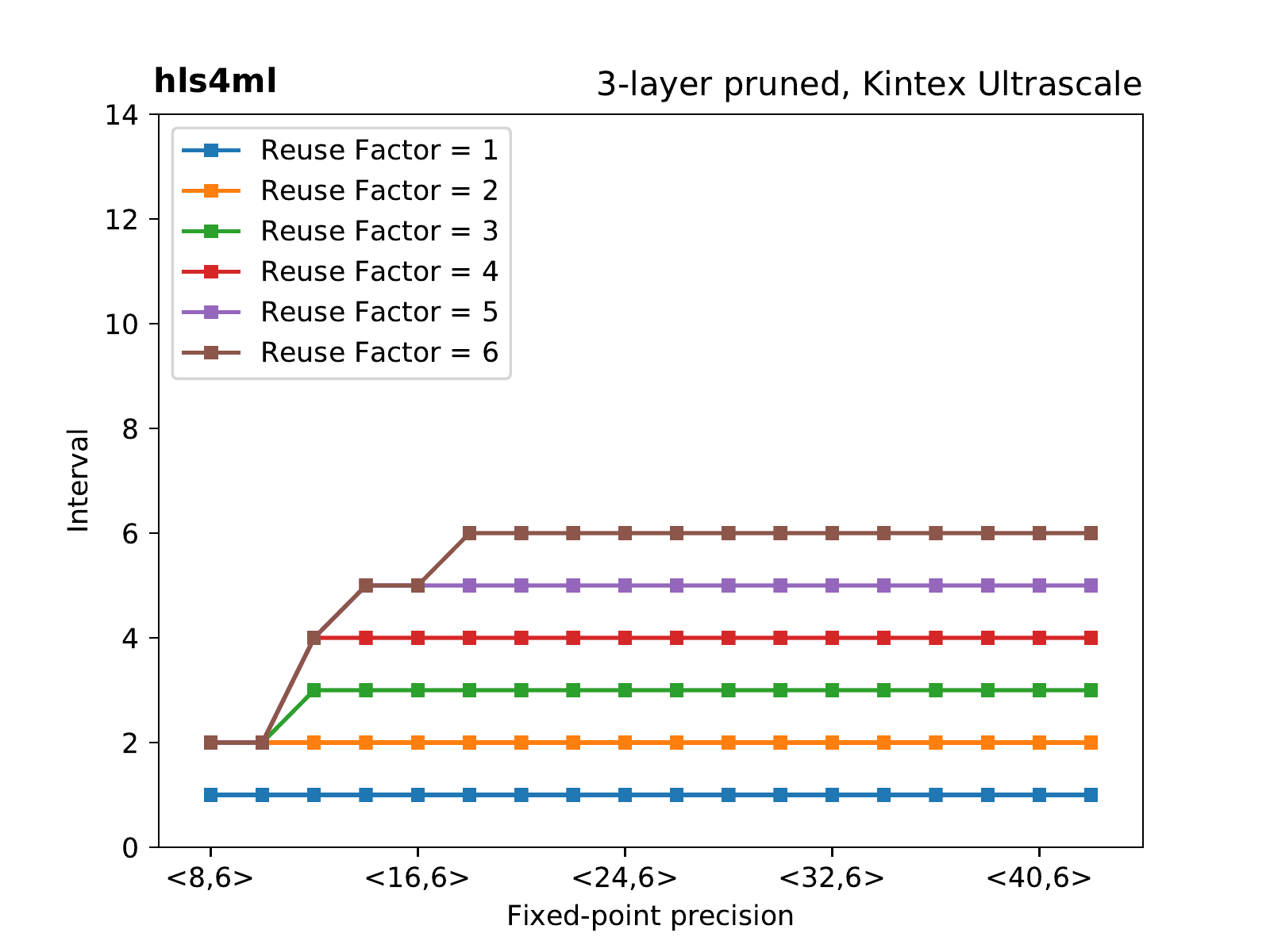}
\end{center}
\caption{Latency (left) and initiation interval (right) in the compressed three-hidden-layer model as a function of the network precision.  The various curves illustrate resource usage for different resource usage factors.  The latency is given in clock cycles for a 200~MHz clock frequency.}
\label{fig:perf-pruned-timing}
\end{figure}

Next, we examine the aspects of the FPGA implemenetation related to latency and throughput.
In Fig.~\ref{fig:perf-pruned-timing}, on the left (right), we show the latency (initiation interval) of the algorithm versus precision for a number of different reuse factors.
The latency of the network inference increases by 4 clocks, corresponding to the four layers of neuron values that must be computed, with each increment in reuse factor.  
This is in line with expectations from Eq.~\ref{eq:latency_one_layer} where additional reuse of multipliers in a given layer calculation incurs added latency.
By design, the initiation interval and the reuse factor match as a new input can be introduced to the algorithm only when all multiplications for a given multiplier are completed.  
At very low network precision, the HLS synthesis initiation interval
is smaller than the reuse factor as multiplications are no longer implemented in DSPs but through FFs and LUTs.

\subsection{FPGA implementation}
\label{sec:implementation}
To get a qualitative understanding of the differences between Vivado HLS resource estimates and a final ``placed and routed'' implementation,
we use a ``bare'' firmware design that uses minimal resources beyond those required by the neural network. 
This ``bare'' implementation consists of a simple VHDL wrapper that connects 
the \hlsfml~firmware block directly to the FPGA's general purpose input/output pins to prevent Vivado from optimizing out the logic we are trying to characterize.
Including the VHDL wrapper, we perform the firmware implementation and compare the resulting resource usage to the Vivado HLS estimates. 

When performing the implementation, we note that the clock period targeted by HLS was not initially achieved due to the design not meeting timing constraints during the implementation step, so we increased the clock period in the final FPGA implementation to meet the timing constraints. 
The required increase in clock period became larger with NN complexity; 
algorithms that took a large part of the FPGA required longer clock periods.
For the three-hidden-layer pruned neural network at 32-bit precision, a clock period of 8~ns was needed to implement an HLS block designed for 5~ns.
This was observed for all reuse factors. 
A simple solution to overcome this issue is to synthesize the HLS design for a slightly smaller clock period than intended.
We also note that different versions of Vivado HLS have varying degrees of success meeting timing constraints.
We had more success meeting timing constraints at the HLS target with Vivado {\tt 2016.4} than {\tt 2017.2}.

\begin{figure}[tbh!]
\begin{center}
\includegraphics[width=0.60\linewidth]{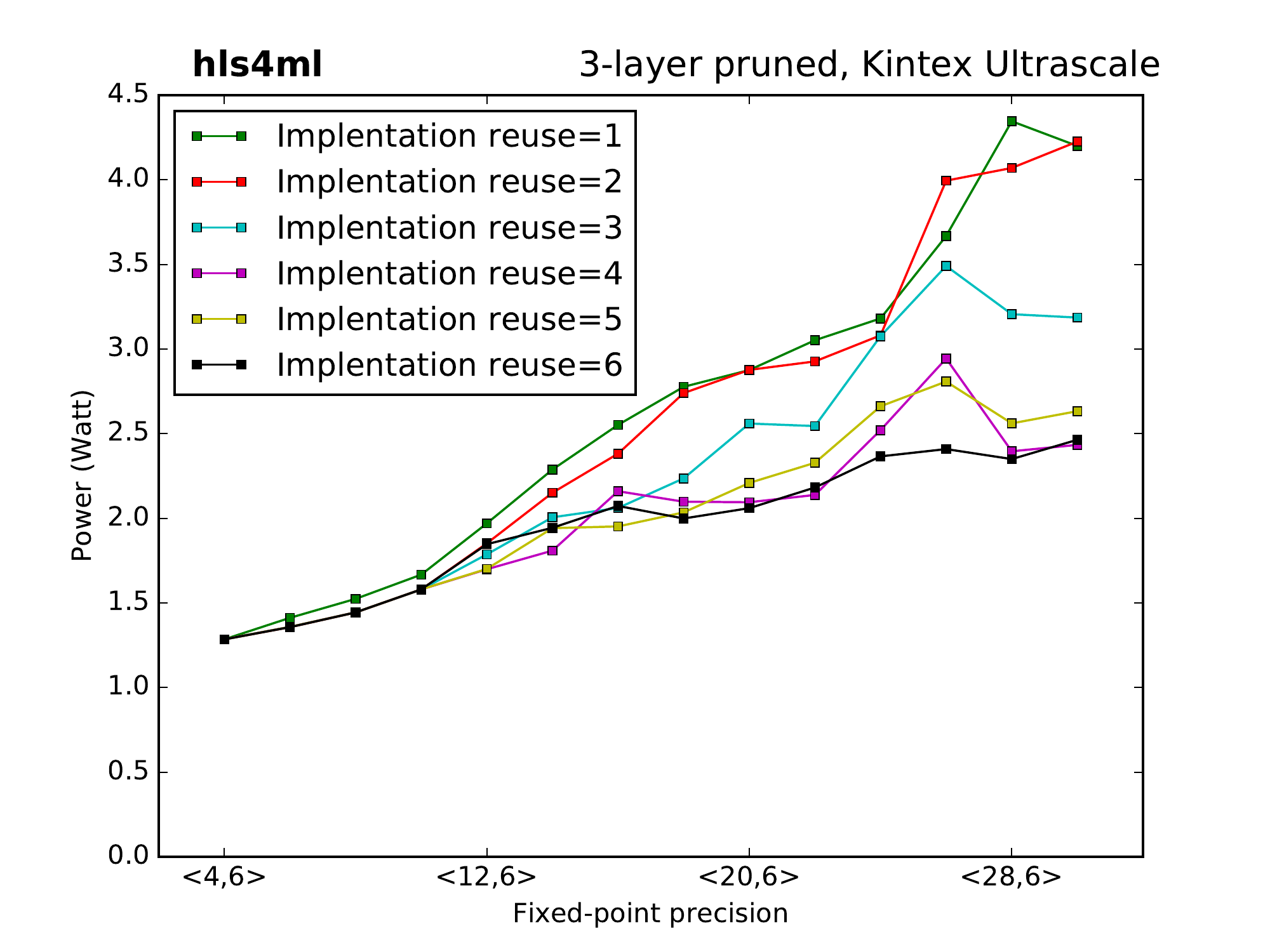}
\end{center}
\caption{Power usage (W) for the baseline three-hidden-layer model as a function of precision }
\label{fig:power-impl}
\end{figure}

The power usage is shown in Fig.~\ref{fig:power-impl}. 
For all implementations, a clear trend towards more power usage for larger network precision is present. 
As expected, as the throughput is decreased by increasing the reuse factor, power usage also goes down.  

%%%%%%%%%%%%%%%%%%%%%%%%%%%%%%%%%%
%% 1 hidden layer implmentation %%
%%%%%%%%%%%%%%%%%%%%%%%%%%%%%%%%%%

We connect each bit of input/output directly to a unique pin 
under the assumption that all inputs are delivered on the same clock edge.
In this ``bare'' implementation, we do not use the high speed transceivers that would allow significantly higher bandwidth input/ouput.
Due to the limited number of pins, we now consider a different neural network model with fewer inputs. 
The model, illustrated in Fig.~\ref{fig:arch} (right), has ten input neurons and one output neuron with one hidden layer between.
We also tested the three-hidden-layer pruned network and find similar quantitative conclusions in the regime where the number of pins was sufficient for implementation.  For the rest of this subsection, we present results with the one-hidden-layer network using an 8 ns clock at implementation.

\begin{figure}[tbh!]
\begin{center}
\includegraphics[width=0.60\linewidth]{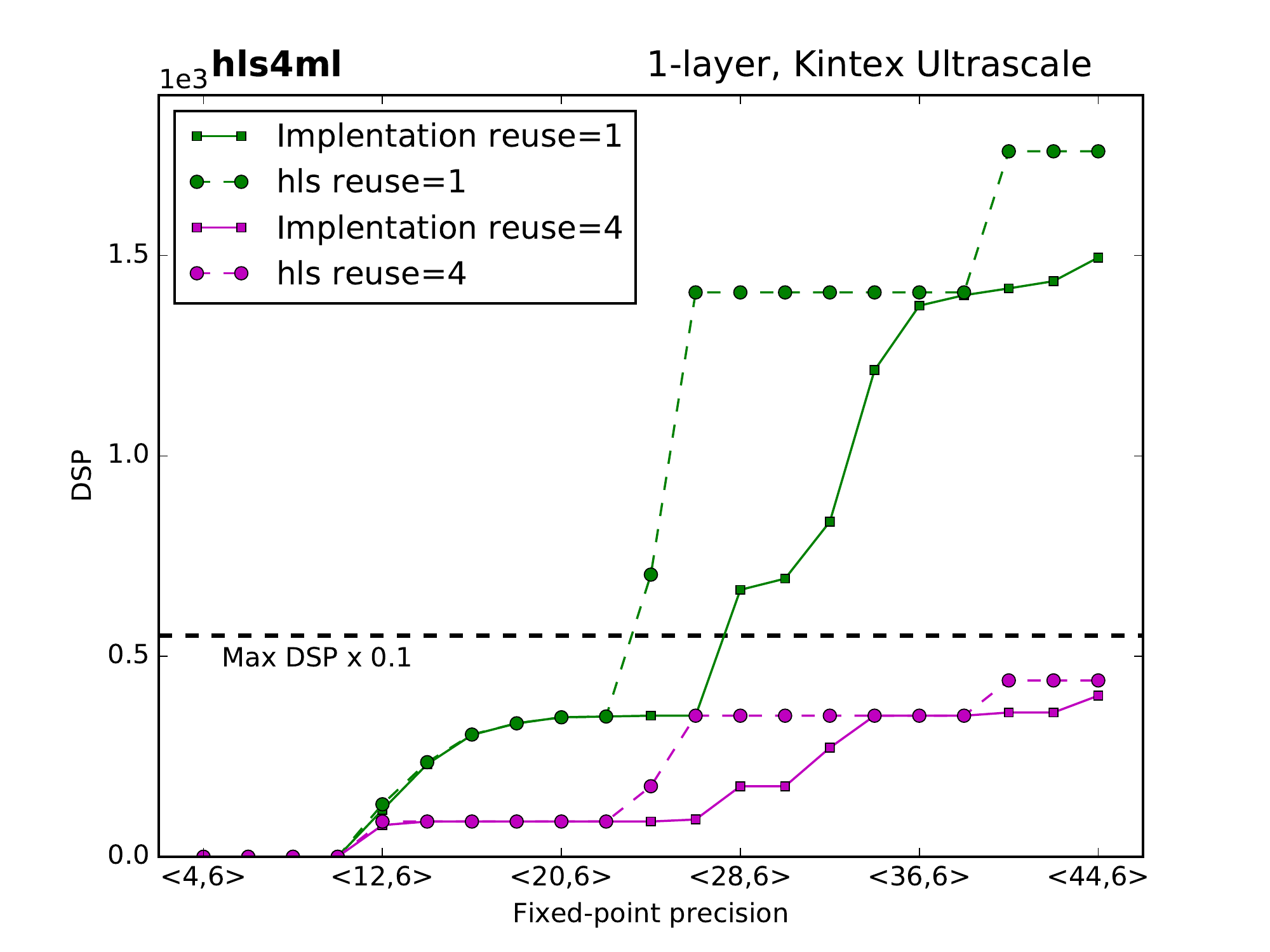}
\end{center}
\caption{Comparsion of the DSP usage of the one-hidden-layer implemenation on the Xilinx Kintex Ultrascale FPGA as a function of the precision for various reuse factors.} 
\label{fig:dsp-impl}
\end{figure}

Figure~\ref{fig:dsp-impl} shows the DSP usage for the implemented design compared with the DSP estimate obtained from the HLS synthesis. 
For all cases, the DSP usage in the implemented design is observed to be smaller than the HLS estimate,
and in particular, we find the HLS synthesis estimate and the implementation agree well for multiplications which require one DSP ($<$ 24 bits).
Deviations between HLS estimates and Vivado implementation are evident above 24 bits, however, as other FF and LUT resources can further be used to optimize DSP usage in the final implementation.

\begin{figure}[tbh!]
\begin{center}
\includegraphics[width=0.48\linewidth]{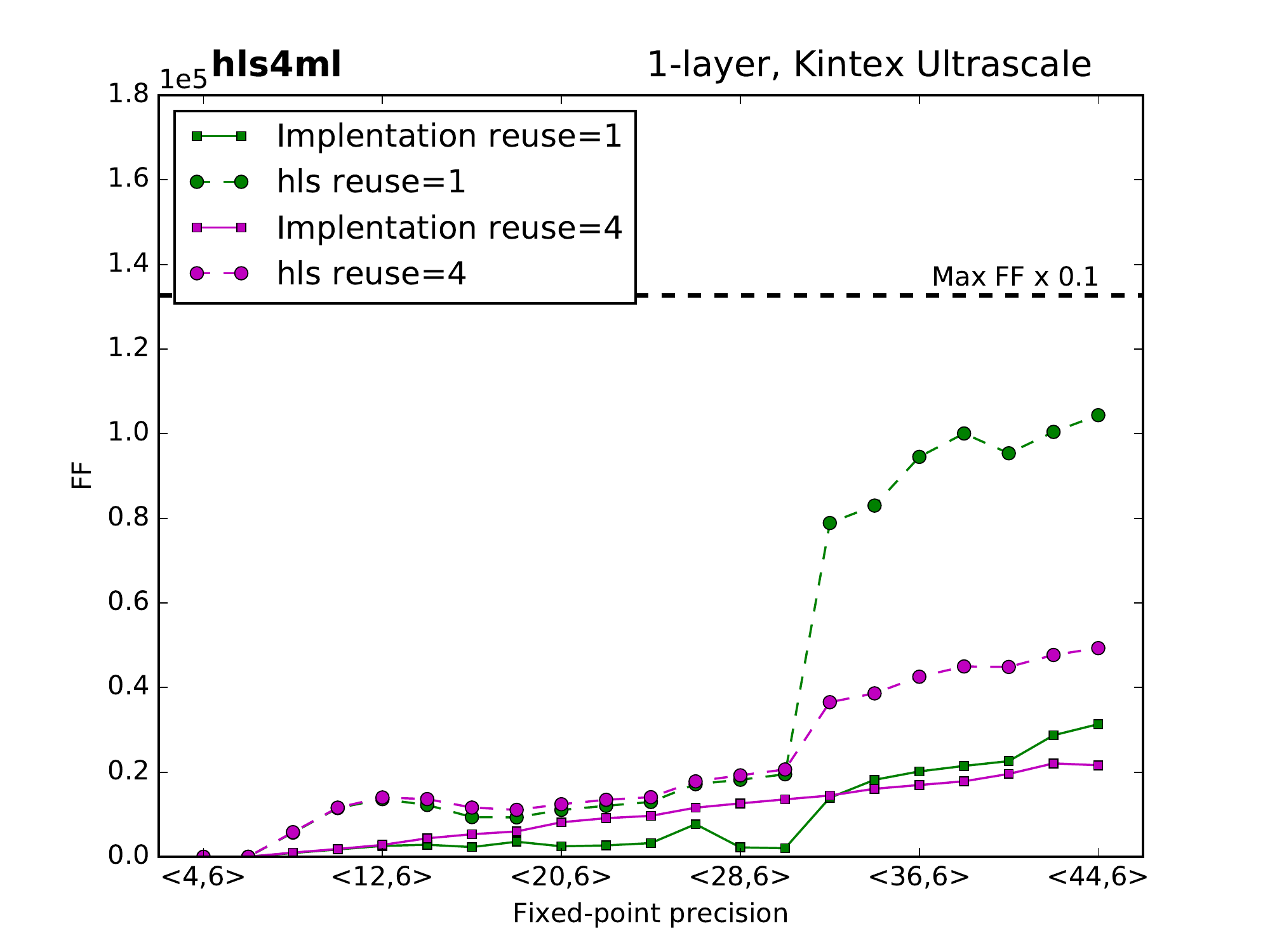}
\includegraphics[width=0.48\linewidth]{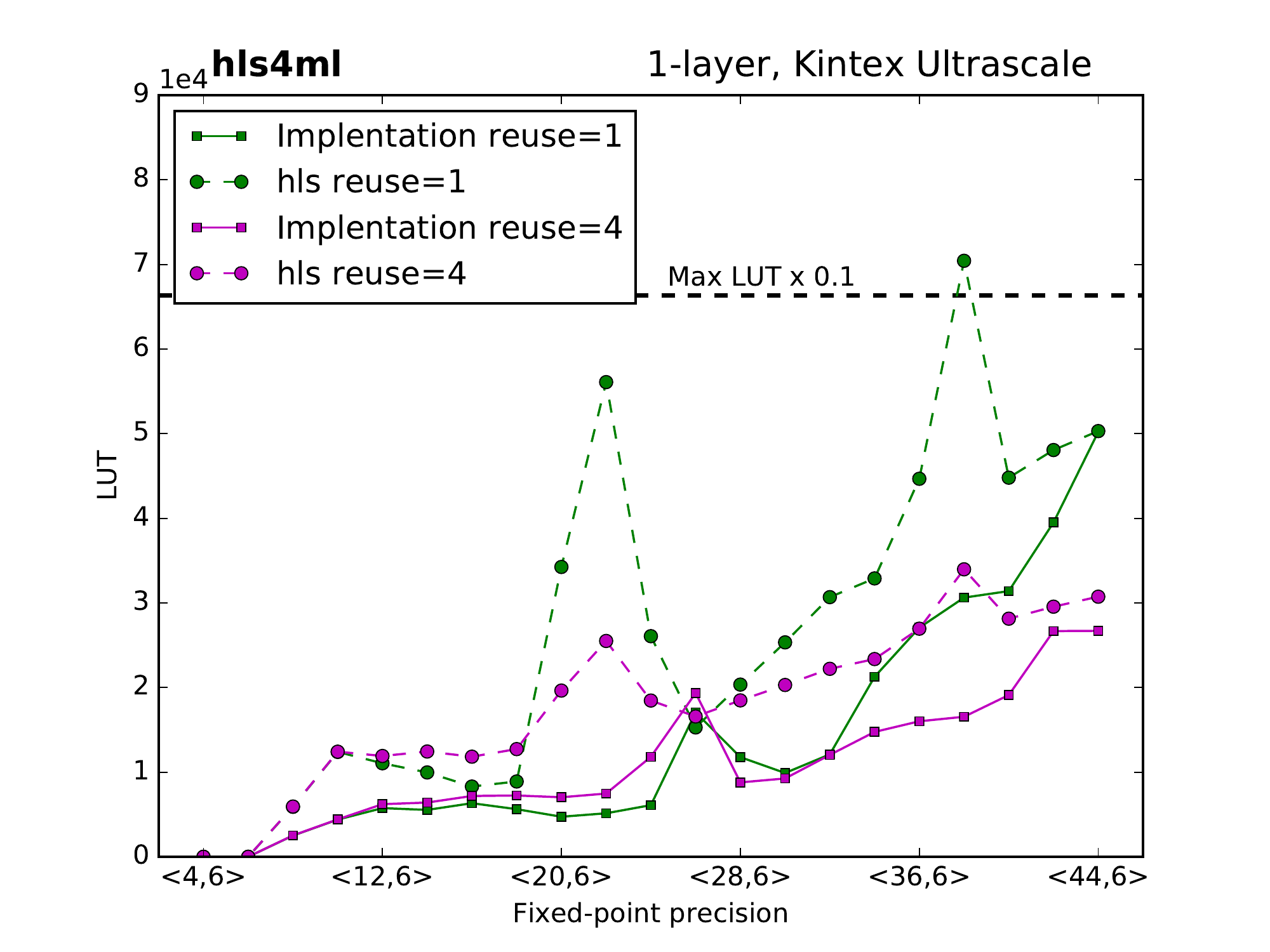}
\end{center}
\caption{Comparison of the FF performance (left) and the LUT performance (right) for the Kintex Ultrascale processor as a function of the precision for 1 and 4 reuse factors.}
\label{fig:lut-ff-impl}
\end{figure}

Figure~\ref{fig:lut-ff-impl} compares the FF and LUT usage between the HLS estimate and the implementation. 
Large differences are present between the HLS estimates and implemented FF usage.
The HLS estimates are typically factors of 2--4 high, becoming particularly large for implementations using more than 32 bits.
The LUT usage is similarly overestimated by the HLS calculation, with the spikes at 22 bits and 38 bits optimized away during the implentation step. 
For all points, excluding the 26-bit implemenation of the LUTs, the HLS estimate is more conservative than the firmware implementation. 

%%%%%%%%%%%%%%%%%%%%%%%%%%%%%%%%%%
%%%%%%%%%%%%%%%%%%%%%%%%%%%%%%%%%%
%%%%%%%%%%%%%%%%%%%%%%%%%%%%%%%%%%

%%%%%%%%%%%%%%%%%%%%%%%%%%%%%%%%%%%%%%%%%%%%%%%%%%%%%%%%%%%%%%%%%%%%%%%%%%%%%%%%%%%%%%%%%%%%%%%%
% S U M M A R Y
%%%%%%%%%%%%%%%%%%%%%%%%%%%%%%%%%%%%%%%%%%%%%%%%%%%%%%%%%%%%%%%%%%%%%%%%%%%%%%%%%%%%%%%%%%%%%%%%
%\clearpage
\section{Summary and Outlook}
\label{sec:outlook}
We introduce \hlsfml, a deep neural network compiler based on HLS capable of porting fully connected networks to an FPGA trained from conventional training frameworks such as {\tt Keras} and {\tt PyTorch}.
For the first result using this framework, we focus on the application of real-time event reconstruction and filtering at the LHC in custom electronics with FPGAs.
This requires pipelined network inference with latencies on the scale of 1~$\mu$s.  
For such low latencies, networks necessarily have a smaller number of parameters. 
For this paper, we consider a specific case study and train a fully connected neural network to identify jets as originating from a light quark, gluon, $W$ boson, $Z$ boson, or top quark.
The original model has $4389$ parameters, and applying network compression and reduced precision, 
it is possible to implement a fully-connected three-hidden-layer network in a Xilinx Kintex Ultrascale
using roughly 10\% of the available DSPs, with results varying with the initiation interval.
The latency of the inference is approximately 75--150~ns with a clock frequency of 200~MHz. 
This fits well into the allowed hardware trigger reconstruction budget of LHC detectors such as ATLAS and CMS. 

The accessibility and ease of configurability in HLS allows for physicists to quickly develop and optimize machine learning algorithms targeting FPGA hardware.
This greatly decreases both firmware development time over traditional VHDL/Verilog-based algorithms and engineering resources in physics which are scarce.
We discuss generic techniques such as network compression, parallelization, and reduced precision, which can be applied to design efficient neural network implementations tailored for different applications at the LHC and beyond.
The results presented use \hlsfml~to scan the network precision and parallelization to optimize DSP and other resource usage.
We compare results of estimated resources from HLS synthesis with the resource usage at implementation.  
HLS resource estimates are conservative, particularly for FFs and LUTs. The HLS resource estimate for DSP usage is comparable to the implemented design although it can be conservative across designed DSP precisions.  

%State about future development
While we have demonstrated the \hlsfml~framework in the context of fully-connected neural networks,
we intend for \hlsfml~to be a general tool for translating many types of neural network architectures that are commonly used in physics. 
For example, we envision expanding the framework to include convolutional neural networks (CNNs) and recurrent neural networks (RNNs), such as Long Short Term Memory (LSTM) units~\cite{LSTM}. 
CNNs are typically used in image based problems including calorimeter cluster reconstruction~\cite{deOliveira:2015xxd,Paganini:2017hrr}. 
LSTMs are used to process a dynamic list of objects; for jet substructure tagging, they have been used to process lists of particle properties belonging to a single jet.
At their core, both network architectures would build on the existing \hlsfml~framework, and the same efficient network design principles apply.
In addition, \hlsfml~can be extended to target FPGAs from other vendors, such as Intel FPGAs using the Quartus HLS compiler. 
We currently support {\tt Keras} and {\tt PyTorch} based implementations.

There are further interesting extensions of this line of research using FPGA co-processors for machine learning, pioneered by efforts such as Micrsoft Brainwave~\cite{brainwave,brainwave2} and others.  
At experiments like CMS and ATLAS, the first tier of reconstruction is limited to the microsecond timescale. 
However, the second tier of reconstruction, the high-level trigger, is limited to reconstruct and understand events on the 100~ms timescale. 
This second tier currently reads the output of the first tier at a rate reduced by a factor of 400 from the original collision rate. 
At these timescales, the inference times can be as long as 10~ms. 
For such long timescales, a large reuse factor can allow for large machine learning algorithms to placed on FPGAs. 
FPGAs can consequently be used as a co-processor accelerator to significantly reduce the time needed to perform complex, core LHC reconstruction algorithms such as track reconstruction.  
With the ability to infer $\mathcal{O}(100)$ times faster than CPUs~\cite{DBLP:journals/corr/HanLMPPHD16}, 
FPGAs can be employed as a low-power, low-cost co-processor in conjunction with CPUs that can be used to significantly speed up the high-level trigger,
and potentially improve its performance. 
In future works, we look forward to more detailed comparisons of neural networks for physics applications on CPU, GPU, and FPGA hardware.

Beyond the LHC, the scope is very broad. 
We believe this tool can be used in many different scientific applications.
Increasingly in nuclear and particle physics, more intense beams and higher rate experiments are being developed, and readout and processing in these experiments often require high speed inference of complex data inputs.

%%%%%%%%%%%%%%%%%%%%%%%%%%%%%%%%%%%%%%%%%%%%%%%%%%%%%%%%%%%%%%%%%%%%%%%%%%%%%%%%%%%%%%%%%%%%%%%%
% Acknowledgements
%%%%%%%%%%%%%%%%%%%%%%%%%%%%%%%%%%%%%%%%%%%%%%%%%%%%%%%%%%%%%%%%%%%%%%%%%%%%%%%%%%%%%%%%%%%%%%%%
\section*{Acknowledgements}
\label{sec:acknowledgements}

We would like to thank Evan Coleman, Marat Freytsis, and Andreas Hinzmann for assistance in producing datasets in a related work.
We thank Jeffrey Berryhill, Doug Burger, Richard Cavanaugh, Eric Chung, Scott Hauck, Andrew
Putnam, Ted Way, and members of Xilinx for useful conversations. 
Neural network training is run on CERN and Amazon AWS GPU resources.
Amazon AWS GPU resources are provided through Fermilab as part of a
DOE "Field Work Proposal", and in particular, we would like to thank
Lothar Bauerdick and Burt Holzmann for their support.
We also thank Xilinx and Ettus Research for sponsoring the 2017 RFNoC and Vivado HLS challenge.
J.~D., B.~K., S.~J., R.~R., and N.~T. are supported by Fermi Research Alliance, LLC under Contract No. DE-AC02-07CH11359 with the U.S. Department of Energy, Office of Science, Office of High Energy Physics.
P.H. is supported by a Massachusetts Institute of Technology University grant. 
M.~P. and J.~N. are supported by the European Research Council (ERC) under the European Union's Horizon 2020 research and innovation program (grant agreement n$^o$ 772369).
Z.~W. is supported by the National Science Foundation under Grants No. 1606321 and 115164.

\bibliographystyle{JHEP}
\bibliography{fml}

\end{document}